\documentclass[12pt,aps,onecolumn,superscriptaddress,showpacs,preprintnumbers,longbibliography]{revtex4-1}
\usepackage[colorlinks,bookmarks=false,citecolor=blue,linkcolor=red,urlcolor=blue]{hyperref}
\usepackage{physics}
\usepackage{colortbl,amsthm,txfonts}
\usepackage{verbatim}
\usepackage{graphicx}
\usepackage{epsfig}
\usepackage{dcolumn}
\usepackage{bm}

\usepackage{amssymb}
\usepackage{amsmath}

\makeatletter

\usepackage{xcolor}
\usepackage{etoolbox}
\makeatletter

\patchcmd{\l@section}{1.0em}{1.5em}{}{}
\patchcmd{\l@subsection}{1.5em}{2em}{}{}
\renewcommand{\@pnumwidth}{2em}
\renewcommand{\@tocrmarg}{3.0em}
\makeatother

\begin{document}
\title[Correlated electrons in flat bands]{Correlated electrons in flat bands: Concepts and Developments}
\author{Madhuparna Karmakar}
\email{madhuparna.k@gmail.com}
\affiliation{Centre of Advanced Computational and Theoretical Sciences, College of Engineering and Technology,\\ 
SRM Institute of Science and Technology, Kattankulathur, Chennai-603203.}
\affiliation{Department of Physics and Nanotechnology, College of Engineering and Technology, \\ 
SRM Institute of Science and Technology, Kattankulathur, Chennai-603203.}

\begin{abstract}
When the electronic dispersion in a material is independent of momentum, it gives rise to strongly correlated flat bands, with the single 
particle energy, quenched. Though the notion of flat bands had been known since long, their experimental realization is achieved 
much later with the advent of ultra cold atomic gases, followed by photonic lattices, coordination polymers and more recently solid 
state materials. By the virtue of their quenched kinetic energy scales the flat band materials provide an ideal platform to engineer, customize 
and investigate the interplay between topology, geometry and strong electronic correlations; giving rise to exotic quantum phases such as, 
unconventional superconductivity, Mott insulator, non Fermi liquid metals etc. This review presents a comprehensive overview of the 
theoretical foundation and material realization of the many body systems with flat electronic bands. We discuss the origin of the flat 
bands and their mathematical construction in prototypical lattices, particularly focussing on those with Lieb and Kagome geometries. 
Observations made and inferences drawn based on the recent experimental and theoretical investigations are presented along with the 
framework for a non perturbative numerical approach to address the quantum phases in the flat band materials. By synthesizing insights 
from both theory and experiment, this review aims to provide a unifying perspective on the emergent many-body phenomena in flat band 
systems and to outline future directions for the field.
\end{abstract}

\date{\today}
\maketitle

\tableofcontents

\section{Introduction}
The electronic properties of any material is innate to its energy dispersion spectra, i. e. the solution of its corresponding Schrödinger equation. 
For a free electron, with the constant rest mass $m_{0}$ in vacuum the electronic wave function is a plane wave $\psi({\bf r}, t) = Ae^{i({\bf k.r}-\omega t)}$ 
with the corresponding solution to the Schrödinger equation being $E({\bf k}) \propto k^{2}$ and the effective electronic mass being $m^{*}/m^{0}$, as 
shown in Figure 1(a). When subjected to a periodic potential in the form of an underlying lattice the electronic wave function needs to be 
defined in terms of the Bloch wave function $\phi({\bf r}, t)$, such that, $\phi({\bf r}, t) = \psi({\bf r}, t)u({\bf r})$, wherein $u({\bf r})$ 
takes into account the lattice periodicity. While a parabolic $E({\bf k}) \propto k^{2}$ dispersion can still be obtained in the limit of weak periodic potential 
of the lattice, the electrons are no longer free and are renormalized as quasiparticles with their low energy excitations being dependent on the 
electron-electron and electron-lattice coupling, thereby recalibrating the effective electronic mass. Bringing the relativistic quantum mechanics in to this 
picture significantly alters the electronic dispersion which now corresponds to the solution of the Dirac equation rather than the 
Schrödinger equation. The corresponding dispersion is linear in momentum, $E({\bf k}) \propto k$, as shown in Figure 1(b), with the 
effective electronic mass $m^{*} \rightarrow 0$. Such linear dispersion bands are known as Dirac bands which intersects each other to give rise to gapless 
energy spectra quantified by the Dirac cones. Though theoretically predicted early on \cite{wallace_pr1947}, the first material realization of Dirac 
bands was in 2005 with the discovery of graphene, which revolutionized the world of two-dimensional (2D) materials and brought into the picture the 
concept and importance of band topology in quantum materials \cite{novoselov_nature2005}. More recently, an even more esoteric electronic dispersion 
was observed in certain class of materials, known as the {\it flat band materials}  \cite{maksymenko_ijmpb2015,sondhi_comprendphys2013,flack_advphysx2018}. As the name suggests, these materials are characterized by momentum independent energy dispersion i. e. $E({\bf k}) \propto k^{0}$, arising out of the destructive interference between the Bloch wave functions, as shown in Figure 
1(c). The corresponding effective electronic mass in such bands is therefore $m^{*} \rightarrow \infty$, since the kinetic energy of the electrons are quenched. The experimental realization of flat band lattices (and materials) such as those with Lieb, Kagome and Moire geometry is fairly recent, both in case of natural and engineered systems \cite{flack_advphysx2018}. Their theoretical predictions however, predates the experimental realization and can be traced back to the pioneering works by Lieb \cite{lieb_prl1989}, Sutherland \cite{sutherland_prb1986}, Tasaki \cite{tasaki_prl1992,tasaki2_epjb2008}, Mielke \cite{mielke_jphysmath1991,mielke2_jphysmath1991}, and others. 
\begin{figure*}
\begin{center}
\includegraphics[height=5.0cm,width=14cm,angle=0]{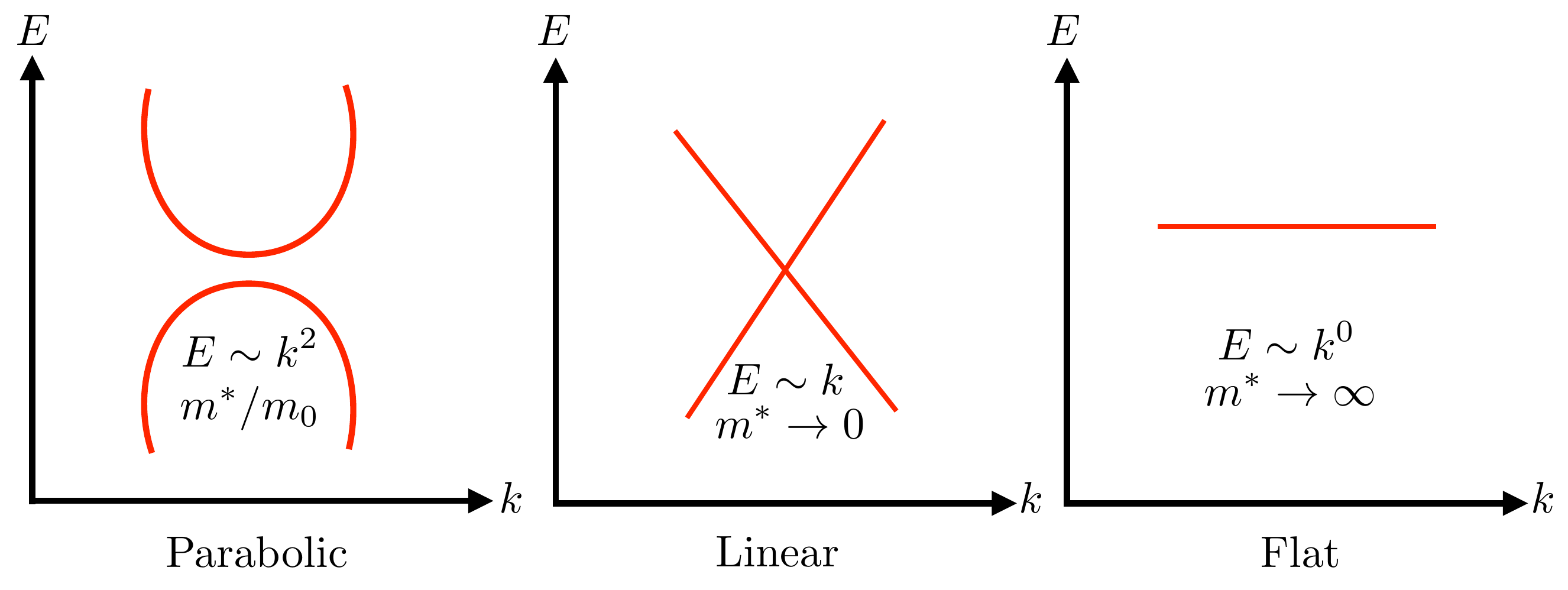}
\label{fig1}
\caption{Schematic band structures illustrating: Parabolic bands with finite effective electronic mass, $k$-linear Dirac bands with 
vanishing effective electronic mass and flat electronic bands with infinite effective electronic mass.}
\end{center}
\end{figure*}

The breakdown of the band theory of solids brought into the focus the physics of strong electronic correlations, particularly in solids with narrow energy bands. 
The origin of novel quantum phases and phase transitions such as, unconventional superconductivity, Mott insulator-metal transition, non Fermi 
liquid (NFL) metal etc. can now be tracked back to such strong electronic correlations in certain classes of materials. In general,  the properties of these 
quantum materials are dictated by the delicate balance between the single particle physics which governs the kinetic energy scale (say, $t$) and the electronic 
interactions (say, $U$) with brings into effect the quantum correlations. As is well known, standard perturbative solutions exist at the two extremes of 
this competition viz. $t \ll U$ and $t \gg U$, while for the regime $t \sim U$ non perturbative approaches are unavoidable \cite{fazekas_book}. Quenching 
the kinetic energy as in case of a flat electronic band thus essentially leaves us with only the correlation scale which now dominates the energy landscape of 
the system such that, even a small $U$ can give rise to significant electronic correlations. In other words, flat bands are inherently strong correlated 
even in the regimes considered to be weak coupling as per the conventional definition of electron-electron interaction \cite{fazekas_book}. This is particularly true in systems comprising either a single isolated flat band or if the gap ($\delta$) between the flat and the dispersive bands is larger than the interaction 
energy scale i. e. $\delta \gg U$ \cite{huhtinen_prb2022}. On the other hand, systems containing band touching points between the flat and the dispersive 
bands are unconventional in their own right as they are effectively multi-band materials hosting both inter and intra band correlations. It is therefore imperative 
that one requires to go beyond the single particle dispersion spectra and electronic band theory to capture the quantum correlations in the realistic flat band 
materials.   

Over the last couple of years a significant amount of work has been carried out to understand the impact of strong electronic correlations in flat band 
materials, both from the perspective of theoretical and numerical foundation as well as material realization. The discovery of graphene led to other 2D 
materials and heterostructure, together known as van der Waals (vdW) materials. Characterized by covalently bonded atoms in layers which are stacked 
together via van der Waals interaction, these materials provided the much required platform for the fundamental understanding of flat band materials as well 
as their potential application as high performance photonic and optoelectronic devices \cite{geim_nature2013,novoselov_science2016,wang_small2020,nelson_natmat2011,baugher_natnanotech2014}. This was quickly followed by the angle dependent stacking of the 2D layers of vdW materials, graphene and other heterostructure leading to incommensurate Moire superlattice structures which 
significantly modified the optical and electronic properties of these materials in terms of carrier mobility, band gap opening and anomalous quantum 
phenomena such as, quantum Hall effect, Hofstadter butterfly etc. \cite{puretzky_acsnano2016,geng_jphyschemlett2020,huang_nanolett2016,wang_acsnano2016,hunt_science2013,dean_nature2013,ponomarenko_nature2013,yankowitz_natphys2012,yang_natmat2013,wang_science2015,giovannetti_prb2007,amet_prl2013,gorbachev_science2014,woods_natphys2014,park_natphys2008,park_prl2008,yu_natphys2014}. The rather unexpected experimental observation of unconventional superconductivity and correlated insulating state in twisted bilayer graphene (TBLG) with Moire lattice structure was the key to draw the attention of the researchers towards the quantum correlations in flat band materials \cite{bristritzer_pnas2011,andrei_natmat2020}. Pristine graphene with its honeycomb lattice structure though topologically non trivial, doesn't host a flat band. On the other hand, Moire structure arising out of the lattice mismatch in TBLG and other heterostructure such as, graphene/hexagonal boron nitride (h-BN) comprise of (nearly) dispersionless electronic bands which promotes correlated electronic phases that can be controlled via changing the twist angle between the layers \cite{liu_science2014}.    

Very recently the interest has shifted over to the class of systems with perfectly flat bands, achieved either via fine tuning of the material parameters 
or due to symmetry protected topology. Primarily, two lattice geometries are in focus viz. Lieb (decorated square) and Kagome, them being line-graph 
lattices are interconvertible via external perturbations. Both these lattices are characterized by a three-site unit cell and are multi-band systems comprising 
of a flat and two Dirac bands. In an ideal Lieb lattice the flat band is tied to the Fermi level and touches the Dirac bands at the $M$-point, while in the Kagome lattice the flat band is at high energy and touches the Dirac bands at the $\Gamma$ point, the two Dirac bands in this lattice intersects each other at the $K$($K^{\prime}$) points. One of the primary reasons for the upsurge in research on these lattice geometries is their possible engineering in artificial set ups, which allows external control over lattice parameters and their response to perturbations. Both Lieb and Kagome lattice geometries have been realized in artificial systems such as, optical  \cite{xing_prb2010,manninen_pra2010,takahashi_sciadv2015,takahashi_natcom2020,stamper_kurn_prl2012} and electronic lattices \cite{drost_natphys2017,slot_natphys2017,du_sciadv2018}, photonic waveguides \cite{baba_natphot2008,yoshino_jpcm2004,atwater_prl2010,aoki_prb2010,nixon_prl2013,kajiwara_prb2016,nakata_prb2012,vicencio_prl2015,mukherjee_prl2015,maczewsky_natcom2017,mukherjee_natcom2017,mukherjee_opticlett2017,real_scirep2017,xia_opticlett2016,zong_opticexp2016,jacqmin_prl2014,whittaker_prl2018,klembt_apl2017,gulevich_prb2016}, metal organic framework (MOF) \cite{jiang_accchemres2021,ni_mathorizon2022,wang_chemsocrev2021} etc. Realizing the perfectly flat band of the ideal Lieb lattice requires significant fine tuning of the parameters and therefore its solid state manifestations are rare, 
the closest being the CuO$_{2}$ planes in the high temperature cuprates, the host of unconventional superconductivity \cite{bednorz_zphysb1986}. In contrast, Kagome lattices are much more well investigated, both in terms of their realization in engineered systems \cite{stamper_kurn_prl2012} as well as in solid state materials \cite{ortiz_prm2019,ortiz_prl2020}.  The geometric frustration of the Kagome lattice provides the ideal platform for stabilizing exotic quantum phases such as, chiral spin liquid \cite{messio_prl2012,capponi_prb2019,bauer_natcom2014,he_prl2014,wietek_prb2015,gong_prb2015,messio_prl2017}, Z$_{2}$ spin liquid \cite{yan_science2011,depenbrock_prl2012,jiang_natphys2012,nishimoto_natcom2013,kolley_prb2015,mei_prb2017,lauchli_prb2019}, valence bond 
solid \cite{marston_jap1991,syromyatnikov_prb2002,nikolic_prb2003,singh_prb2007,budnik_prl2004,evenbly_prl2004,schwandt_prb2010,poiblanc_prb2010,poiblanc_prb2011} etc., as the ground state of Kagome quantum magnets. Further, the recent discovery of Kagome metals have opened up the possibilities for the interplay between strong electronic correlations and geometric frustration, bringing forth phases like, spatially modulated superconductivity \cite{chen_nature2021,schwemmer_prb2024}, cascade of charge order \cite{zhao_nature2021,korshunov_prb2025}, nematic order \cite{nie_nature2022}, NFL metal \cite{checkelsky_natphys2024},  flat band induced gapless insulator \cite{shashi_kagome2025} etc.

Keeping in pace with the experimental efforts the interplay of strong electronic correlations and flat electronic bands in Lieb and Kagome lattices 
have been investigated using a battery of analytic and semianalytic approaches \cite{keisel_prl2013,weber_prb2018,wang2_prb2013,dora_prb2011,zhao_prl2006,weeks_prb2010}. One of the most important concept posit by these 
studies is the geometric weight and quantum metric, particularly in the context of superconductivity in flat band systems \cite{peotta_natcom2015,liang_prb2017,liang2_prb2017,liang_pra2019,julku_prl2021,huhtinen_prb2022,pentilla_comphys2025,jiang_pnas2024,julku_prl2016}. 
A quenched kinetic energy effectively means a vanishingly small superfluid weight/phase stiffness and therefore an absence of superconductivity. It was 
however observed that even in case of isolated flat bands there exists a finite superfluid weight and therefore a nonzero superconductivity. Based on extensive theoretical investigations it is inferred that in flat band lattices the superfluid weight comprises of two parts viz. the conventional weight (${\cal D}_{conv}$) and 
a geometric weight (${\cal D}_{geom}$) \cite{huhtinen_prb2022}. While the former, dependent on the electronic dispersion vanishes for a flat band,  the later dictated by the quantum geometry of the Bloch states, survive. More explicitly, the quantum geometric tensor depends on the intrinsic geometry of the Bloch 
states and not just on the dispersion $\epsilon_{\bf k}$. It encodes how the Bloch state $\vert u_{n{\bf k}}\rangle$ changes as one moves through the momentum space, and therefore accounts for the flatness and twist of the electronic wave functions in the momentum space as well as the overlap between the neighboring Bloch states. The real part of the quantum geometric tensor gives the quantum metric which is the measure of the distance between the nearby quantum states in the Hilbert space \cite{huhtinen_prb2022}.

Sophisticated non perturbative numerical techniques such as, determinant quantum Monte Carlo (DQMC)  \cite{scalettar_prb2014,zhu_prr2023,costa_prb2017,li_prb2022,costa_prb2016}, dynamical mean field theory (DMFT) 
and its variants \cite{julku_prl2016, huhtinen_prbl2021,kumar_prb2017}, static path approximated (SPA) quantum Monte Carlo \cite{swain_prr2020}, functional renormalization group (FRG) \cite{wu_prl2021,schwemmer_prb2024}, variational cluster approximation (VCA) \cite{yu_prb2012}, density matrix renormalization group (DMRG) \cite{zhu_prbl2021} and several others have been used to solve the many body Hamiltonian (such as, Hubbard model) in flat band systems. As expected, this has proved to be particularly challenging in case of the Kagome metals owing to their highly frustrated geometry which restricts the formation of (quasi) long range order and leads to severe fermionic sign problem. However, these studies have brought forth interesting effects of the Kagome flat band on the quantum phases in such systems. It was observed that the flat band aids in the transient localization of itinerant fermions at and close to the Fermi level  giving rise to a gapless insulating phase, a {\it la} Anderson sans potential disorder \cite{shashi_kagome2025}. 

Quantum materials with flat electronic bands is a rather new and extremely promising area with many of its intriguing features yet to be unveiled. 
This article overviews the current status of this field sketching out the important experimental and theoretical works, keeping primarily in focus the flat band 
materials with Kagome and Lieb geometry. While every attempt has been made to accommodate and highlight the important literature, with the field being continuously expanding the bibliography is nowhere complete and I apologize to those being left out. The rest of the article is structured as follows:  
Section II gives an overview of the flat band lattices in terms of their realization in engineered platforms and solid state materials, this is followed by section 
III which discusses the mathematical construction of certain prototypical lattices with flat electronic bands, based on generic tight binding model Hamiltonian. 
The quantum phases and phase transitions in flat band systems arising out of electronic correlations is presented in section IV, while section V outlines  
the methodological details of a non perturbative numerical approach which can handle strong electronic correlations in flat band materials. We conclude 
with section VI by presenting a future outlook to this field.  

\section{Flat band lattices: engineered and natural}
The early theoretical proposals of flat electronic band dates back to 1980's when Sutherland proposed the existence of ``strictly localized states'' in Dice lattice \cite{sutherland_prb1986}. This was followed by the seminal work by Lieb on the Hubbard model proving that certain bipartite lattices (such as, the decorated square lattice) with flat bands exhibit macroscopic ferrimagnetism at half filling \cite{lieb_prl1989}. Independent works on ``flat band ferromagnetism'' by Mielke and Tasaki soon followed and it was observed that a macroscopic ferromagnetic state can be established when the lowest energy band is flat \cite{mielke_jphysmath1991,mielke2_jphysmath1991,tasaki_prl1992,tasaki2_epjb2008}. It was realized that owing to the macroscopic degeneracy of the flat band, it is possible to construct the ``compact localized state'' (CLS) which are Wannier-like eigenstates, from the linear combination of degenerate Bloch states.  Based on their response to the applied magnetic field these flat band systems could further be classified into (a) non-chiral and (b) chiral. In the former the applied magnetic field could destabilize the flat band via destroying its origin in the fine tuned interference, for example in Mielke's and Tasaki's lattices. The chiral flat band lattices such as, dice and Lieb, on the other hand are protected by sub-lattice symmetry, i. e. the edge (rim) sites which gives rise to the flat band are connected only to the corner sites which lies at the origin of the Dirac bands \cite{aoki_prl1993,ohberg_prb2012,xiao_annphys2017,matsumura_prb1996}. Such symmetry protected lattices do not depend on the magnitude of the 
coupling strength and are governed by the local topology of the site connectivity. The chiral flat bands therefore remain macroscopically degenerate to a great 
extent even in presence of an applied magnetic field. Figure 2 illustrates the structures of some of the common flat band lattices such as, Kagome, 
Lieb, dice, Creutz ladder, and diamond chain. In each case the unit cell comprises of more than one atom leading to a multi-band dispersion spectra.  
\begin{figure*}
\begin{center}
\includegraphics[height=5.0cm,width=15cm,angle=0]{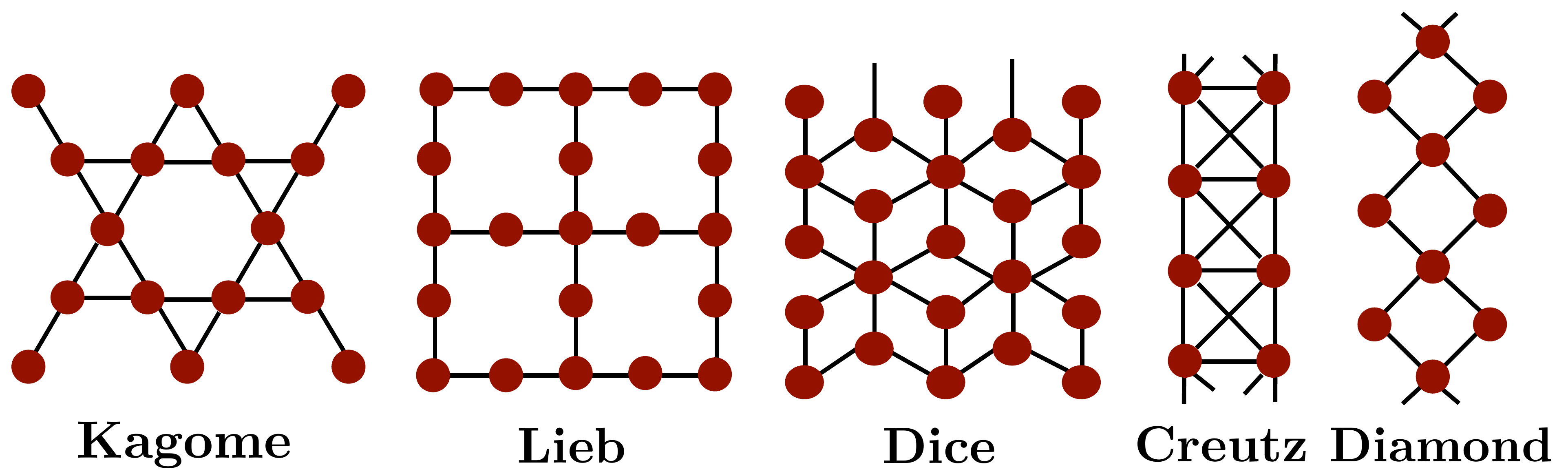}
\label{fig2}
\caption{Representative structures of some flat band lattices: (from left to right) Kagome, Lieb, Dice, Creutz ladder and Diamond chain.}
\end{center}
\end{figure*}

\subsection{Optical lattices}
In spite of these early theoretical proposals the experimental realization of flat band lattices took a significantly long time owing to the technological challenges 
and it was only around 2010, with the advent of ultracold atomic gas setups that the theoretical proposals could achieve experimental realization. The first experimental proposal for an optical Lieb lattice was made by Shen {\it et al.} in 2010, based on six detuned standing wave laser beams. The depths of the sub-lattices were controlled by the relative amplitude of the laser beams \cite{xing_prb2010}. It was observed that the band structure of shallow or moderately deep sub-lattice deviates from that of an ideal Lieb lattice and the flat band gains a finite width. Apaja {\it et al.} showed via the numerical computation of the exact Bloch wave spectrum that the relative amplitudes of the laser beams could be tuned to achieve an almost perfectly flat band with width only 1.5$\%$ of the total band width \cite{manninen_pra2010}. Further, simulating the wave packet dynamics of this system showed that while the fermionic cold atoms remain confined in the flat bands the repulsively interacting bosons tend to tunnel and escape through the dispersive bands. Similar observations were later made in the context of other quasi one-dimensional (1D) systems, as well \cite{chien_jphysb2016,manninen_pra2013}. 

For a bosonic system the first experimental realization of optical Lieb lattice was reported by the group of Takahashi in 2015, who used five standing wave 
lasers to produce a sufficiently deep lattice hosting a flat band \cite{takahashi_sciadv2015}. A dynamic tuning of the optical lattice was carried out in this experiment to transfer the bosonic condensate from the ground state to the flat band via a two step process. The experimental observations were found to be in agreement with the predictions based on the numerical simulations and the bosonic condensate was found to undergo interband tunnelling and decay into the lower dispersive band due to interactions \cite{manninen_pra2010}. The lifetime of the flat band was therefore dependent on the decay of the condensate; the opening up of a gap between the flat and the dispersive bands reduces the tunnelling and thus could potentially increase the flat band lifetime \cite{manninen_pra2010,takahashi_sciadv2015}. In 2017 the first experimental realization of the fermionic cold atoms in the Lieb lattice was carried out by Taie {\it et al.} who demonstrated the spatial adiabatic passage of the fermions, akin to the stimulated Raman adiabatic passage technique used for highly efficient transfer of atomic internal states. In the fermionic optical Lieb lattice experiment by Taie {\it et al.} dark state-mediated transfer of fermions was observed between the corner and the rim sites \cite{takahashi_natcom2020}.  The first realization of the Kagome geometry in the optical lattice setup was carried out via overlaying two commensurate triangular optical lattices generated by lights with different wave length \cite{stamper_kurn_prl2012}. However, since the flat band in the Kagome lattice constitute the highest energy state, probing it via cold atoms is challenging.  

\subsection{Electronic manipulation}
Advancement in lithography and atomic manipulation techniques provide an alternate avenue for the prospective nanoscale engineering of the flat band lattices. Using scanning tunnelling microscopy (STM) 2D Lieb lattice was embedded on a substrate surface by Liljeroth and co workers \cite{drost_natphys2017} and independently by Swart and co workers \cite{slot_natphys2017}. In the former, atoms were removed from chlorine monolayer to construct the Lieb lattice \cite{drost_natphys2017} while in the later carbon monoxide molecules were added to the substrate giving rise to a Lieb lattice structured repulsive potential which the surface electrons were subjected to \cite{slot_natphys2017}. In these set ups the spatially resolved electron density could be measured using conductance spectroscopy while the Bloch waves corresponding to the flat and dispersive bands could be selectively accessed by changing the bias voltage. Electronic Kagome lattice was designed on the surface of twisted multilayer silicene using STM and scanning tunnelling spectroscopy (STS), by Li {\it et al.}\cite{du_sciadv2018}. It was noted that the origin of the Kagome geometry and the associated flat band was due to the twisting of the silicon (Si) layers which gave rise to a screened Coulomb periodic potential which the free electrons were subjected to. 

\subsection{Photonic lattices and waveguides}
In photonics, flat bands are intimately connected to the technologically significant concept of slow light, wherein the reduction of group velocity enhances 
nonlinear effects and enables efficient pulse buffering \cite{baba_natphot2008}. Early proposals for achieving photonic flat bands were based on photonic 
crystal slabs of high-index dielectric rods \cite{yoshino_jpcm2004}. Experimental investigations of photonic flat bands received the much needed impetus 
in 2010 with the proposals for obtaining non-interacting single particle bands in plasmonic waveguide networks \cite{atwater_prl2010,aoki_prb2010},  followed 
by the realization of geometric frustration in a setup comprising of coupled laser arrays arranged in the Kagome lattice geometry \cite{nixon_prl2013}.  
Further, flat bands in Lieb \cite{kajiwara_prb2016} and Kagome \cite{nakata_prb2012} lattices were obtained using terahertz spoof plasmons, femtosecond 
laser written waveguide arrays \cite{vicencio_prl2015,mukherjee_prl2015},  periodic driving \cite{maczewsky_natcom2017,mukherjee_natcom2017,mukherjee_opticlett2017} , optical logic gates based on CLS \cite{real_scirep2017}, to name a few. 
In a similar spirit, experimental realization of flat band lattices were carried out using optically induced lattices by applying laser writing beams to a photo refractive 
medium. Optically induced Lieb \cite{xia_opticlett2016} and Kagome \cite{zong_opticexp2016} lattices were created by using the superposition of mutually 
coherent square lattices and with a single induction beam, respectively. Finally, flat band lattices were also engineered using polariton condensates, where the observations of flat bands were confirmed based on the photoluminescence spectra \cite{jacqmin_prl2014,whittaker_prl2018,klembt_apl2017,gulevich_prb2016}. 

\subsection{Organic polymers}
A relatively recent and arguably one of the most active areas of research on engineered flat band lattices are the molecular framework materials such as, conjugated polymers (CP), metal organic framework (MOF) and the covalent organic framework (COF)\cite{jiang_accchemres2021,ni_mathorizon2022,wang_chemsocrev2021}. These organic materials comprising of metal ions conjugated to organic ligand molecules as in MOFs or organic molecules linked via covalent bonds as in COFs are chemically tuneable porous systems \cite{liu_jamchmesoc2020,yuan_jamchemsoc2020,yuan_natcom2018} which allows for their applications in gas storage \cite{furukawa_science2013,murray_chemsocrev2009}, separation \cite{kim_science2020,bloch_science2012}, catalysis \cite{yang_chemsocrev2017,jiao_advmat2018}, proton conduction \cite{wang_chem2017}, sensors \cite{schneemann_chemsocrev2014,kreno_chemrev2012} etc. The conventional MOFs are however, insulators that limits their application as multifunctional electronic devices. On the other hand the 2D MOFs are typically semiconductors and owing to the extended $\pi$-conjugation in their 2D planes,  promote delocalization of the charge carriers within the network 
leading to high mobility and conductivity \cite{hmadeh_chemmat2012,sun_angewcheminted2016}. Along with the existing positive aspects of the conventional (3D) MOFs the 2D MOFs are characterized by high stability, electrochemical activity, photoactivity, customizable band gaps, high electrical conductivity, magnetic order, topological states \cite{ko_chemcommun2018,meng_chemrev2019,xie_chemrev2020,koo_chem2019} etc., that opens up the avenue for their applications in MOFtronics \cite{campbell_angcheminted2015,wu_jamchemsoc2017,arora_advmar2020,huang_natcom2015,dong_natcom2018,yang_natcom2019,li_chemsci2017,park_jamchemsoc2018,wada_angcheminted2018,sheberla_natmat2017,feng_natenergy2018}. Amongst the several possibilities, three particular 2D MOF and COF 
structures have stood out viz. Lieb, Kagome, honeycomb lattices and their combinations which continues to be extensively investigated in the context of 2D functional materials \cite{jiang_accchemres2021,ni_mathorizon2022,wang_chemsocrev2021}. Among the synthesized MOFs (and COFs) some of those with Lieb lattice geometry includes Fe-phthalocyanine MOF (FePc-MOF) \cite{oi_advphysres2025}, sp$^{2}$C COF \cite{cui_natcom2020,jiang_natcom2019} etc. 
In a similar spirit MOFs and COFs with a Kagome lattice geometry have been synthesized in M$_{3}$C$_{6}$O$_{6}$ network deposited on Ag(111) substrate \cite{shaiek_advmatintf2022}, Fe(II) ions assembled on Ag(111) substrate \cite{hua_jphyschemlett2021}, Cu$_{3}$(C$_{6}$S$_{6}$) (Cu‑benzenehexathiolate) i. e. Cu-BHT \cite{takenaka_sciadv2021}, to name a few. Finally, DCBP$_{3}$CO$_{2}$ MOF synthesized on a G/Ir(111) surface comprises of a honeycomb lattice geometry \cite{kumar_nanolett2018}. 

\subsection{Solid state materials}
Solid state materials with ideal Lieb lattice structure are difficult to find and require significant fine tuning of the parameters to obtain a perfectly flat band at the Fermi level. Most of the existing realization of the Lieb lattice are therefore in engineered materials, as discussed above. However, one of the most prominent solid state manifestation of the Lieb lattice geometry is in the CuO$_{2}$ planes of the cuprates, a feature shared by all the members of the cuprate family, wherein each copper atom is surrounded by six oxygen atoms \cite{bednorz_zphysb1986}.

In contrast to the Lieb lattice, solid state materials with the geometrically frustrated Kagome lattice structure have been widely investigated, particularly 
in the context of spin-$1/2$ quantum magnets, bringing forth intriguing signatures such as, pinch points \cite{moessner_prb1998,henley_annrevcondmatphys2010} observed via neutron scattering experiments on spin ice \cite{fennell_science2009}, Dirac spin liquid \cite{zeng_prbl2022,ma_prl2009}, valence bond crystal \cite{matan_natphys2010} and other quantum spin liquids for example,  in herbertsmithite compounds \cite{helton_prl2007,helton_prl2010,khuntia_natphys2020}. More recently the discovery of Mn$_{3}$Sn \cite{nakatsuji_nature2015,nayak_sciadv2016,kuroda_natmat2017,kimata_nature2019,collington_natcom2019,wuttke_prb2019}, Fe$_{3}$Sn \cite{fenner_jpcm2009,hou_advmat2017,kang_nature2018,yin_nature2018,lin_prl2018,li_apl2019,li_prl2019,tanaka_prb2020}, 
Co$_{3}$Sn$_{2}$S$_{2}$ \cite{liu_natphys2018,wang_natcom2018,yin_natphys2019,liu_science2019,shen_apl2019,lachman_natcom2020}, Gd$_{3}$Ru$_{4}$Al$_{12}$ \cite{nakamura_prb2018,matsumura_jpsj2019}, FeSn \cite{inoue_apl2019,kang_natmat2019,lin_prb2020,sales_prm2019} and the AV$_{3}$Sb$_{5}$ family (with A=K, Rb, Cs) \cite{ortiz_prm2019,ortiz_prl2020} have brought into the focus the Kagome metals, governed by strong electronic correlations; unlike their strongly coupled insulating counterpart the Kagome metals do not have any spin analogue. These materials 
are reported to host novel quantum phases such as, time-reversal symmetry (TRS) broken charge ordering (CO) \cite{mielke_nature2022,yang_sciadv2020,labollita_prb2021,yu_prb2021,khasanov_prr2022}, orbital nematic order \cite{nie_nature2022,asaba_natphys2024,li_natcom2022}, cascade of CO with hierarchies of ordering wave vectors \cite{zhao_nature2021,korshunov_prb2025}, unconventional superconductivity \cite{ortiz_prl2020,ortiz_prm2019,tazai_sciadv2022,guguchia_natcom2023,zheng_nature2022,chen_prl2021,wang_prr2023,du_prb2021}, chiral loop current order \cite{tazai_pnas2024} etc. Particularly intriguing is the AV$_{3}$Sb$_{5}$ family which is reported to exhibit an unconventional CO with T$_{co} \sim $ 90K along with the onset of an anomalous Hall effect. Importantly, the onset of this CO is accompanied by TRS breaking without any signature of local moment formation \cite{mielke_nature2022,yang_sciadv2020,neupert_natphys2022,yu_prb2021,khasanov_prr2022}. Possible nematic transition and/or a one-dimensional CO is reported at T$_{nem} \sim$ 30-50K, sensitive to external perturbations such as, strain or an applied magnetic field \cite{nie_nature2022,asaba_natphys2024,li_natcom2022}. Further, transition to an unconventional superconducting phase is observed at T$_{sc} \sim $ 1K in these vanadium based Kagome metals \cite{ortiz_prl2020,ortiz_prm2019,tazai_sciadv2022,guguchia_natcom2023,zheng_nature2022,chen_prl2021,wang_prr2023,du_prb2021}. 

In a similar spirit, $x$-ray diffraction on FeGe showed a dimerization driven short range charge density wave (CDW) transition at T$_{CDW} \sim $ 105K, whose microscopic origin continues to be debated. Further, a pressure induced transition to a long ranged $\sqrt{3}\times \sqrt{3}$ CDW order from the short ranged $2\times 2$ CDW order is observed above a pressure of $\sim$ 15GPa, in FeGe. Both the short and long range CDW orders coexist over the pressure range $4 < p \lesssim 12$ GPa \cite{korshunov_arxiv2025}. A NFL phase arising out of flat band induced localization of the itinerant fermions akin to that of heavy fermion metals was reported in Ni$_{3}$In \cite{checkelsky_natphys2024}, which was further supported by the observed sublinear temperature dependence of the longitudinal resistivity in this material \cite{han_prbl2024}. A relatively less explored Kagome material is LaRu$_{3}$Si$_{2}$, comprising of Kagome layers of Ru sandwiched between layers of La and Si with honeycomb lattice structure. LaRu$_{3}$Si$_{2}$ is a fully gapped superconductor with T$_{c} \sim$ 7K, the highest among the Kagome superconductors, attributed to the flat band, the van Hove point being close to the Fermi level and the high density of states from the narrow Kagome bands \cite{mielke_prm2021}. Moreover, a cascade of CO was observed at high temperatures in La(Ru$_{1-x}$Fe$_{x}$)$_{3}$Si$_{2}$ ($x=0-0.05$) giving rise to a CO transition at T$_{CO} \sim $ 400K, i. e. a CO at room temperature \cite{plokhikh_comphys2024}.   

\section{Mathematical construct of flat band lattices}
In this section we discuss the origin and construction of the model Hamiltonian simulating the flat band lattice geometries. 
Figure 3 shows the lattice geometry and the corresponding single particle dispersion spectra for the Lieb and Kagome lattices, the prototypical 
flat band systems. Although the honeycomb lattice doesn't host a flat band we discuss it in the same footing as the Kagome and Lieb because, (i) it is important from the point of view of graphene and therefore Moire lattices (comprising of flat band) and (ii) it is a line-graph lattice to Kagome just as the Lieb and Kagome lattices are to each other, and are therefore interconvertible by tuning a single lattice parameter. Such an inter-convertibility is particularly useful in case of engineered flat band lattices such as, MOFs and COFs, so as to bring forth materials with customized properties via tuning through external perturbations. We begin by discussing the generic tight binding (TB) Hamiltonian in 2D, which reads as \cite{jiang_accchemres2021}, 
\begin{figure*}
\begin{center}
\includegraphics[height=6.0cm,width=14cm,angle=0]{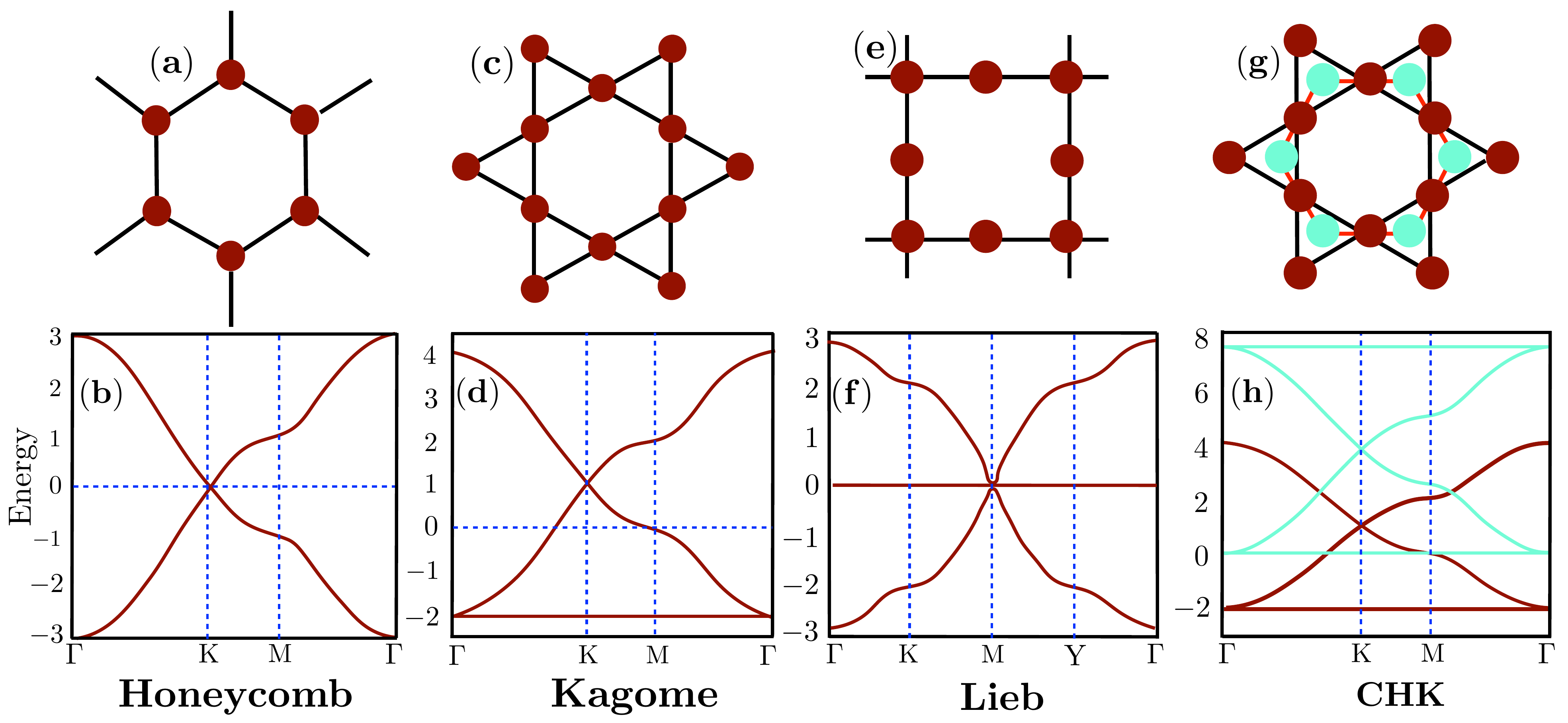}
\label{fig3}
\caption{Illustration of lattices (top panels) and their corresponding tight binding electronic band structure (bottom panels). 
From left to right: honeycomb, Kagome, Lieb and combined honeycomb-Kagome (CHK) lattice.}
\end{center}
\end{figure*}

\begin{eqnarray}
H & =& \sum_{i\alpha} \epsilon_{i\alpha}c_{i\alpha}^{\dagger}c_{i\alpha}-(\sum_{\langle i\alpha, j\beta\rangle}tc_{i\alpha}^{\dagger}c_{j\beta} + 
\sum_{\langle \langle i\alpha, j\beta \rangle \rangle}t_{2}c_{i\alpha}^{\dagger}c_{j\beta} + \sum_{\langle \langle \langle i\alpha, j\beta \rangle \rangle \rangle} 
t_{3}c_{i\alpha}^{\dagger}c_{j\beta}) + i\lambda \sum_{\langle \langle i\alpha, j\beta \rangle \rangle} c_{i\alpha}^{\dagger}\sigma_{z}(\frac{\vec d_{jk}}{\vert \vec d_{jk}\vert} \times \frac{\vec d_{ki}}{\vert \vec d_{ki}\vert})c_{j\beta} \nonumber \\ && + M\sum_{i\alpha}c_{i\alpha}^{\dagger} \sigma_{z}c_{i\alpha}
\end{eqnarray} 
where, $\epsilon_{i\alpha}$ is the on-site energy of the $\alpha$-orbital at the $i$th site and $c$'s and $c^{\dagger}$'s are the electron annihilation and creation operators. The nearest (NN), next-nearest (2NN) and third next nearest (3NN) neighboring lattice sites are denoted as, $\langle i, j\rangle$, $\langle \langle i, j\rangle\rangle$, $\langle \langle \langle i, j\rangle\rangle \rangle$, respectively with the corresponding hopping integrals being $t$, $t_{2}$ and $t_{3}$. The strength of the 2NN intrinsic spin-orbit coupling (SOC) is denoted by $\lambda$ while $\sigma_{z}$ is the $z$-component of the Pauli matrix. The last two terms of the Hamiltonian correspond to the intrinsic SOC and Zeeman exchange field, respectively. In the momentum space the Hamiltonian is expressed as, 
\begin{eqnarray}
H(\vec {k}) & = & 
\begin{bmatrix}
\epsilon & H_{0} \\
H_{0}^{*} & \epsilon
\end{bmatrix}
\end{eqnarray}
with $H_{0}$ encoding the hopping integrals corresponding to the different lattice types. 

\subsection{Honeycomb lattice geometry}
Figure 3(a) shows the structure of a honeycomb lattice, characterized by a two-site unit cell with the electronic band structure comprising 
of two Dirac bands intersecting each other at the $K$ and $K^{\prime}$-points, giving rise to Dirac points, as shown in Figure 3(b). 
The corresponding TB Hamiltonian reads as, 
\begin{eqnarray}
H(\vec {k}) & = & 
\begin{bmatrix}
\epsilon & -t(e^{ik_{1}}+e^{ik_{2}}+e^{ik_{3}}) \\
-t(e^{-ik_{1}}+e^{-ik_{2}}+e^{-ik_{3}}) & \epsilon
\end{bmatrix}
\end{eqnarray}
where, $k_{n}$ is defined as $k_{n} = \vec k . \vec a_{n}$ with the NN hopping vectors being $\vec a_{1} = (\frac{-\sqrt{3}}{2}\hat x -\frac{1}{2}\hat y)$, 
$\vec a_{2} = \hat y$ and $\vec a_{3} = (\frac{\sqrt{3}}{2}\hat x -\frac{1}{2}\hat y)$. Inclusion of SOC opens up a non trivial gap at the Dirac point, leading 
to the transition from a semimetal to a topological insulator (TI), with opposite Chern numbers in the two Dirac bands. The first theoretical predictions of MOF 
with honeycomb lattice structure were the organic TI, Bi$_{2}$(C$_{6}$H$_{4}$)$_{3}$ and Pb$_{2}$(C$_{6}$H$_{4}$)$_{3}$ \cite{wang_natcom2013}, with Bi and Pb ions located at the honeycomb lattice sites; this was quickly followed by several other theoretical predictions based on replacing the metal ions with Pd \cite{wang_natcom2013}, In \cite{liu_prl2013}, Tl \cite{su_apl2018}, Mn \cite{wang_prl2013,hu_nanoscalereslett2014}, Fe \cite{kim_prb2016}, V \cite{hu_nanoscalereslett2014} etc. The experimental realization of honeycomb organic framework is rather recent wherein a COF, theoretically predicted to 
be Dirac semimetal was synthesized by triangulenes \cite{palicek_natnanotech2017}. Apart from such single-orbital honeycomb MOFs multiorbital MOFs with honeycomb structure were theoretically predicted in In$_{2}$(C$_{6}$H$_{4}$)$_{3}$ \cite{liu_prl2013}, Tl$_{2}$(C$_{6}$H$_{4}$)$_{3}$ \cite{su_apl2018} etc. which hosts SOC tuned topologically nontrivial gaps and a fractional Chern insulator state in presence of a partially filled flat band.  
 
\subsection{Kagome lattice geometry} 
The Kagome lattice and the corresponding electronic band structure are shown in Figure 3(c) and (d), respectively. With its three-site unit cell 
the Kagome lattice comprises of two Dirac bands which intersect each other at the $K$($K^{\prime}$)-point, along with a high energy flat band, which 
touches the Dirac band at the $\Gamma$-point. In the momentum space the TB Hamiltonian for the Kagome lattice reads as, 
\begin{eqnarray}
H(\vec {k}) & = & 
\begin{bmatrix}
\epsilon & -2t\cos k_{3} & -2t\cos k_{2} \\ 
-2t\cos k_{3} & \epsilon & -2t\cos k_{1} \\ 
-2t\cos k_{2} & -2t\cos k_{1} & \epsilon
\end{bmatrix}
\end{eqnarray}
Inclusion of SOC opens up a gap at the $K$($K^{\prime}$)-points and at the band touching $\Gamma$-point giving rise to a TI, with the flat and the bottom 
Dirac bands having nonzero spin Chern number with opposite sign and the middle band has a zero spin Chern number.  The first theoretical prediction for the 
MOFs/COFs with the Kagome geometry was that of quantum spin Hall (QSH) effect in Ni$_{3}$(C$_{6}$S$_{6}$)$_{2}$ \cite{wang_nanolett2013}, 
comprising of three Ni ions located at the Kagome sites. The design protocol for this material was subsequently used to identify and engineer topological characteristics in several other MOFs such as, metal-dicyanoanthracene \cite{zhang_nanolett2016}, HTT-Pt \cite{zhang_prb2016}, 
anitalo-based lattice \cite{ni_nanoscale2018} etc. Later, ferromagnetism was experimentally observed in the Kagome MOF Cu(1, 3-bdc) along with possible topological magnon state \cite{chisnell_prl2015}. Kagome geometry is also realized in coloring triangle (CT) lattice, as in Cu$_{2}$(C$_{8}$N$_{2}$H$_{4}$)$_{3}$ exhibiting quantum anomalous Hall (QAH) insulator phase \cite{gao_nanores2020}, diatomic Kagome lattice comprising of Ying-Yang Kagome bands 
was observed in anilato-based MOF Al$_{2}$(C$_{6}$O$_{4}$Cl$_{2}$)$_{3}$ and exhibits topological properties \cite{ni_pccp2020}. 

\subsection{Lieb lattice geometry}
Figure 3(e) and (f) shows the geometry and the non interacting electronic band structure for the Lieb lattice, an edge centerd square lattice 
comprising of two Dirac bands which touches the flat band tied to the Fermi level, at the $M$-point. The corresponding TB Hamiltonian is defined as, 
\begin{eqnarray}
H(\vec {k}) & = & 
\begin{bmatrix}
\epsilon_{c} & -2t\cos(\vec{k. v_{1}}) & -2t\cos(\vec{k.v_{2}}) \\ 
-2t\cos(\vec{k. v_{1}}) & \epsilon_{E} & -4t_{2}\cos(\vec{k.v_{1}})(\vec{k.v_{2}}) \\ 
-2t\cos(\vec{k.v_{2}}) & -4t_{2}\cos(\vec{k.v_{1}})(\vec{k.v_{2}}) & \epsilon_{E}
\end{bmatrix}
\end{eqnarray}
where, $\vec{v_{1}} = \hat x$ and $\vec{v_{2}} = \hat y$, respectively. Note that the lattice contains different on-site energies for the edge-center  
($\epsilon_{E}$) and corner ($\epsilon_{c} = \epsilon_{E} + \Delta E$) sites, $t$ and $t_{2}$ correspond to the NN hopping between the corner and the edge center sites and the 2NN hopping between the edge-center sites, respectively. $\Delta E \neq 0$ opens up an energy gap at the $M$-point either below 
or above the flat band, depending on the sign of $\Delta E$, while a dimerization $t \rightarrow t \pm \delta$ opens up two gaps above and below the flat 
band, finally $t_{2} \neq 0$ induces dispersion in the flat band. The ideal Lieb lattice structure is defined by $\epsilon_{E}=\epsilon_{c}$ and $t_{2}=0$ and 
the perfect flat band can survive only for an ideal Lieb lattice. The first experimental evidence of MOFs/COFs with (a slightly strained) Lieb geometry was in a $sp^{2}$ carbon conjugated COF, an assembly of pyrene (Py) and 1, 4-bis(cyanostyryl) benzene, Py(BCSB)$_{2}$ which was found to support ferromagnetism upon hole doping \cite{jiang_natcom2019,cui_natcom2020}. 
In a similar spirit, intrinsic TI properties were demonstrated in phthalocyanine-based MOFs (MPc-MOFs) with Lieb lattice geometry \cite{jiang_nanolett2020}. Other topological states such as, QSH and QAH were reported in this material, realized via the strain tuned closure of the gap at the Fermi level \cite{jiang_nanolett2020}. Based on the density functional theory (DFT) and tight binding model calculations strain controlled phase transitions between a 
trivial insulator and a TI was predicted for the MPc-MOF with Lieb lattice geometry. Further, it was shown that the inclusion of SOC in this material allows for a transition between the TRS-broken QSH state and a Chern insulating state \cite{jiang_nanolett2020}. In a similar spirit, Stoner ferromagnetic instability was demonstrated in a $sp^{2}$ carbon conjugated COF with dimerized Lieb lattice structure, based on tight binding and first principle calculations \cite{jiang_natcom2019}. 

\subsection{Combined Honeycomb and Kagome lattice geometry}
Honeycomb-Kagome-Lieb are inter convertible line-graph lattices, which allows for possible combinations between them, such as, 
combined honeycomb-kagome (CHK) lattice, shown in Figure 3(g) with the corresponding electronic band structure presented in Figure 3(h). 
Such lattices are fairly common in MOFs and COFs where the two sub-lattices contains different organic ligands and/or metal ions. In presence of weak 
inter-sub-lattice coupling either of the sub-lattices are populated depending upon the electron filling, while strongly coupled sub-lattices promote complex electronic and magnetic properties. CHK lattices are reported in carbon nitride COF, C$_{9}$N$_{4}$ wherein the molecular orbitals formed by the N and C atoms reside on the honeycomb and the Kagome lattices, respectively and the corresponding Dirac and Kagome bands are attributed to the same \cite{chen_jmatchem2018}. The crossing between these bands give rise to nodal ring at the Fermi level, as has been observed in C$_{9}$N$_{4}$, and is topologically protected by the C$_{2}$ rotational symmetry. An intriguing interplay of localized spins and itinerant electrons can be envisaged in the CHK MOFs comprising of $d$-electrons in the Kagome and $\pi$-electrons in the honeycomb sub-lattice, with the spin-degenerate ($p_{x}$, $p_{y}$) honeycomb bands located in between the spin-split Kagome bands. Such an interplay was experimentally realized in Cu-hexaiminobenzene ([Cu(HAB)$_{2}$]), with HAB ligands and Cu ions forming the honeycomb and the Kagome lattices, respectively \cite{jiang_nanoscale2019}. While the $\pi$-electrons of the HAB prefers a ferromagnetic order an antiferromagnetic order is promoted by the $d^{9}$-electrons of the Cu$^{2+}$ ions. Interplay of the localized and itinerant spin degrees of freedom is well investigated in the context of Kondo and heavy fermion materials, bringing forth non trivial material characteristics such as, giant magnetoresistance (GMR) and unconventional superconductivity \cite{anderson_pr1961,kondo_progtheorphys1964}.  

The inter-convertibility of the line-graph Lieb/Kagome lattices was made use of for the purpose of strain controlled band structure reconstruction, it was demonstrated that an applied shear strain is a suitable tuning parameter to continuously deform the Lieb lattice and reconstruct it with a Kagome geometry \cite{jiang_prb2019}. The band structure reconstruction process involves topological phase transition, along with emergent van Hove singularities and type-I, II and III Dirac cones, classified based on their angle of tilt \cite{lang_pra2023}. The protocol was further proposed to be practically realizable using photonic waveguides. Further, strain controlled evolution of the Kagome lattice in presence of SOC showed the transition between semi-metallic and topological phases \cite{ulloa_2dmat2023}. 

\section{Quantum phases and transitions in flat bands}
Having their kinetic energy scales quenched the flat band materials provide the ideal ground to foster correlated electronic phases.  
The diverging density of states at the flat band amplifies the correlation effects, enabling instabilities toward various symmetry-breaking ground states, 
such as, flat band ferromagnetism, unconventional superconductivity, Mott insulating states, NFL metal, topological phases etc. This section focuses 
in particular on the superconducting, Mott insulating and NFL phases of the flat band materials. For flat band ferromagnetism the readers are guided to the 
seminal work by Lieb \cite{lieb_prl1989} while the interplay between the topology and many body correlations in these systems can be best understood from 
the excellent review articles by Parameswaran {\it et al.} \cite{sondhi_comprendphys2013} and Bergholtz {\it et al.} \cite{bergholtz_ijmb2013}.

\subsection{Superconductivity and superfluidity in flat band lattices}
Originally initiated with the discovery of superconductivity in Moire lattices such as, TBLG \cite{cao_nature2018} there has been a recent flurry of research on 
the superconducting properties of strongly correlated systems with flat electronic bands \cite{peotta_natcom2015,liang_prb2017,liang2_prb2017,liang_pra2019,julku_prl2021,huhtinen_prb2022,pentilla_comphys2025,jiang_pnas2024,julku_prl2016}.  Superconductivity in dispersive 2D materials is associated with two important scales, viz. (i) the Cooper pair formation scale, which according to the Bardeen-Cooper-Schrieffer (BCS) theory is given as $T_{c} \propto \exp(-\frac{1}{\vert U\vert \rho_{0}(E_{F})})$, where $\vert U\vert$ is the strength of the attractive interaction and $\rho_{0}(E_{F})$ is the density of states at the Fermi level and (ii) the Berezinskii-Kosterlitz-Thouless (BKT) transition scale quantifying the 
onset of macroscopic phase coherence. For a flat band tied to the Fermi level, as in the Lieb lattice the density of states diverges, leading to $T_{c} \propto \vert U\vert$ indicating that the BCS critical temperature for pair formation is much higher in the flat bands as compared to the dispersive bands; an observation which holds particularly true for the isolated flat bands \cite{heikkila_jeptlett2011,khodel_zphys1990,kopnin_prb2011}.  On the other hand the requirement of dissipation less electronic transport in a superconductor brings in the concept of superfluid weight $D_{s}$ or superfluid stiffness, a quantity which determines the BKT transition temperature for macroscopic phase coherence. The standard definition of superfluid weight reads as, $D_{s}=n_{e}/m^{*}$, with $n_{e}$ being the total particle density and $m^{*}$ the effective electronic mass. A quenched kinetic energy as in case of a flat band leads to the localization of the electrons and therefore the divergence of $m^{*}$. In other words, the superfluid weight can be naively expected to vanish in an isolated flat band resulting in the loss of superconductivity.  Unlike single band, in multi-band systems apart from the conventional contribution there is a non-zero geometric contribution to the superfluid weight which survives even in the flat band \cite{peotta_natcom2015,liang_prb2017,julku_prl2016}. The geometric contribution for the flat band is related to the quantum 
metric \cite{provost_communmathphys1980,resta_epj2011,ozawa_prb2018}. Further, it has been demonstrated that for multi-band superconductors the superfluidity of the system is dependent not just on the quantum geometry but also on the type of band touching \cite{huhtinen_prb2022}. Based on the calculations of optical spectral weight at strong coupling, tight lower and upper bounds were determined for the $D_{s}$ and $T_{c}$ using the geometry of the flat band Wannier functions, such that a higher $D_{s}$ and $T_{c}$ is realized for the flat band superconductivity if the energy gap ($E_{0}$) separating the 
flat band from the dispersive bands is small. However, if $E_{0} \ll \vert U\vert$,  interaction scale dominates and the multi-band effects of the system plays significant role in deciding the $T_{c}$ \cite{huhtinen_prb2022}. The effect of quantum metric is of particular importance in the weak coupling regime accessible via the MFT, where the criteria for isolated flat bands is satisfied. 

Figure 4 shows the BKT transition temperature obtained by solving the attractive Hubbard model on the Lieb lattice using MFT. The separation between the 
flat and the dispersive bands in the lattice is quantified in terms of the energy gap ($E_{gap}$) controlled by the staggered hopping $\delta$ between the inter and intra unit cell lattice points, such that $\delta=0$ corresponds to the gap closure at the linear band touching $M$-point. The observed higher T$_{BKT}$ 
for $\delta=0$ as compared to the $\delta \neq 0$ isolated flat band scenario suggests that the participation of multiple energy bands in fact promotes superconductivity, making its realization in real materials, plausible. The transition temperature T$_{BKT}$ is suppressed as the flat band gets progressively isolated from the dispersive bands. Further, for the interaction $\vert U\vert \rightarrow 0$ there is finite superfluid weight in case of the Lieb lattice unlike the square lattice for which T$_{BKT}$ is exponentially suppressed in the weak coupling regime \cite{huhtinen_prb2022}.  
\begin{figure}
\begin{center}
\includegraphics[height=6.5cm,width=6.5cm,angle=0]{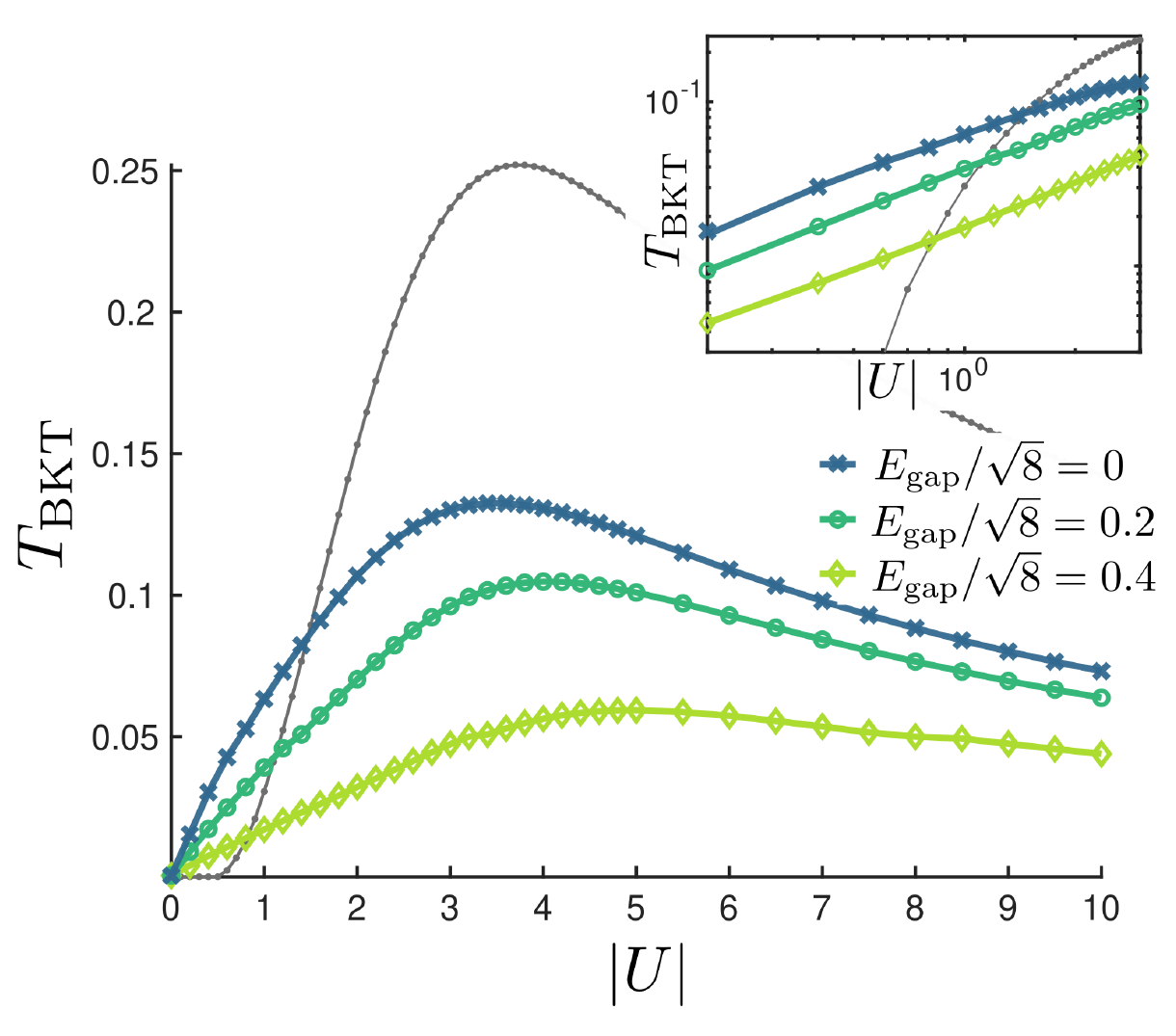}
\label{fig4}
\caption{BKT temperature computed for the square lattice (gray) and for the Lieb lattice with a half-filled flat band (blue, green and yellow) for different 
values of energy gap ($E_{gap}$) between the flat and the dispersive bands. The highest BKT temperature is obtained for $E_{gap}=0$ corresponding 
to the linear band touching at the $M$-point. For the square lattice $T_{BKT}$ is exponentially suppressed in the weak coupling regime whereas $T_{BKT}$ 
on the isolated flat band is proportional to $\vert U\vert$,  for $\vert U\vert \rightarrow 0$. Inset: BKT temperature at interactions $0.2 \le \vert U \vert \le 3$. Figure adapted with permission 
from \cite{huhtinen_prb2022}. Copyright 2022, American Physical Society.}
\end{center}
\end{figure}

Figure 5 presents a comprehensive picture of the various facets of superconductivity and the related transitions on the Lieb lattice as obtained by solving the attractive Hubbard model using different numerical approaches. Figure 5(a) shows the sub-lattice resolved superconducting order parameter ($\Delta_{A}$ and $\Delta_{B}$) as a function of the staggered hopping parameter $\delta$, at different temperatures for a lattice filing of $\nu=1.5$ (half filled Lieb lattice) at an interaction of $U=-0.4J$, where $J$ is the kinetic energy scale and $J=1$ sets the reference energy scale of the system. The results are obtained using DMFT calculations and shows that the superconducting order parameter is significantly larger in the flat band (A-site) as compared to that in the dispersive bands (B-site). In comparison, the results obtained via MFT calculations show that the correct $\delta$-dependence of the order parameter can't be captured by the MFT \cite{julku_prl2016}. DQMC calculations on the Lieb lattice showed non trivial interplay between the charge transfer and superconducting pair correlation, such that at the lattice filling of $\rho=2/3$ and $4/3$, corresponding to the fermionic densities at which the flat band is occupied for the first time and is completely filled, respectively, the pair correlation undergoes suppression, while for $\rho=1$ and $5/3$ the pair structure factor ($P_{s}$) is maximum \cite{scalettar_prb2014}. This is shown in Figure 5(b) at the selected interactions of $U=-4t$ and $U=-8t$, where $t=1$ sets the reference energy scale of the system.  Further, MFT calculations were used to study the population imbalanced superconductivity in a Lieb lattice. As a function of the chemical potential and an applied in-plane Zeeman field, finite momentum superconductivity in the form of Fulde-Ferrell-Larkin-Ovchinnikov (FFLO) and $\eta$-phases, arising out of inter and intraband pairing involving the flat band were found to be stabilized, as shown in Figure 5(c). It was noted that unlike the conventional finite momentum pairing dictated by the shift in the Fermi surface corresponding to the minority component, for a flat band lattice the finite momentum paired state arises due to the complete deformation of the Fermi surface corresponding to one of the pairing components \cite{tylutki_prb2018}. The influence of the quantum metric on the flat band superfluidity/superconductivity was studied using DMFT on a Lieb lattice and it was observed that the contribution to the superfluid weight from the flat band is twice as large as compared to that of the dispersive bands \cite{julku_prl2016}. As shown in Figure 5(d) at the selected lattice filling of $\nu=1.5$ and $\nu=2.5$, the geometric superfluid weight $D_{s,geom}$ (red) far exceeds the conventional contribution $D_{s,conv}$ (blue) across the interaction regime. Moreover, it was demonstrated based on DMFT that in the weak coupling regime ($\vert U\vert \sim t$) there is an insulator-pseudogap crossover at finite temperatures, wherein the insulating phase is attributed to the flat band \cite{huhtinen_prbl2021}. The physics of BCS-BEC crossover on the Lieb lattice was investigated based on the SPA quantum Monte Carlo technique. Using the spectroscopic and thermodynamic properties of this system the thermal transition and crossover scales were mapped out, with a significant pseuodogap regime being stabilized at intermediate interaction and temperature, as shown in Figure 5(e) \cite{swain_prr2020}. SPA was further utilized to bring forth an alternate protocol for superconductor-insulator transition (SIT) in the Lieb lattice, viz. the applied strain. Modelled in terms of lattice dimerization corresponding to asymmetric inter and intra unit cell hopping integrals on a Lieb lattice the applied strain was shown to be a cleaner alternative to the potential disorder induced SIT. Increasing strain was found to result in band structure reconstruction and suppression 
of the (quasi) long range phase coherence leading to the progressive loss of superconductivity, as observed in Figure 5(f). By mapping out the finite temperature phases and the thermal scales, an experimentally realizable protocol for SIT was put forward in this work \cite{swain_prr2020}.   
\begin{figure*}
\begin{center}
\includegraphics[height=7.0cm,width=15cm,angle=0]{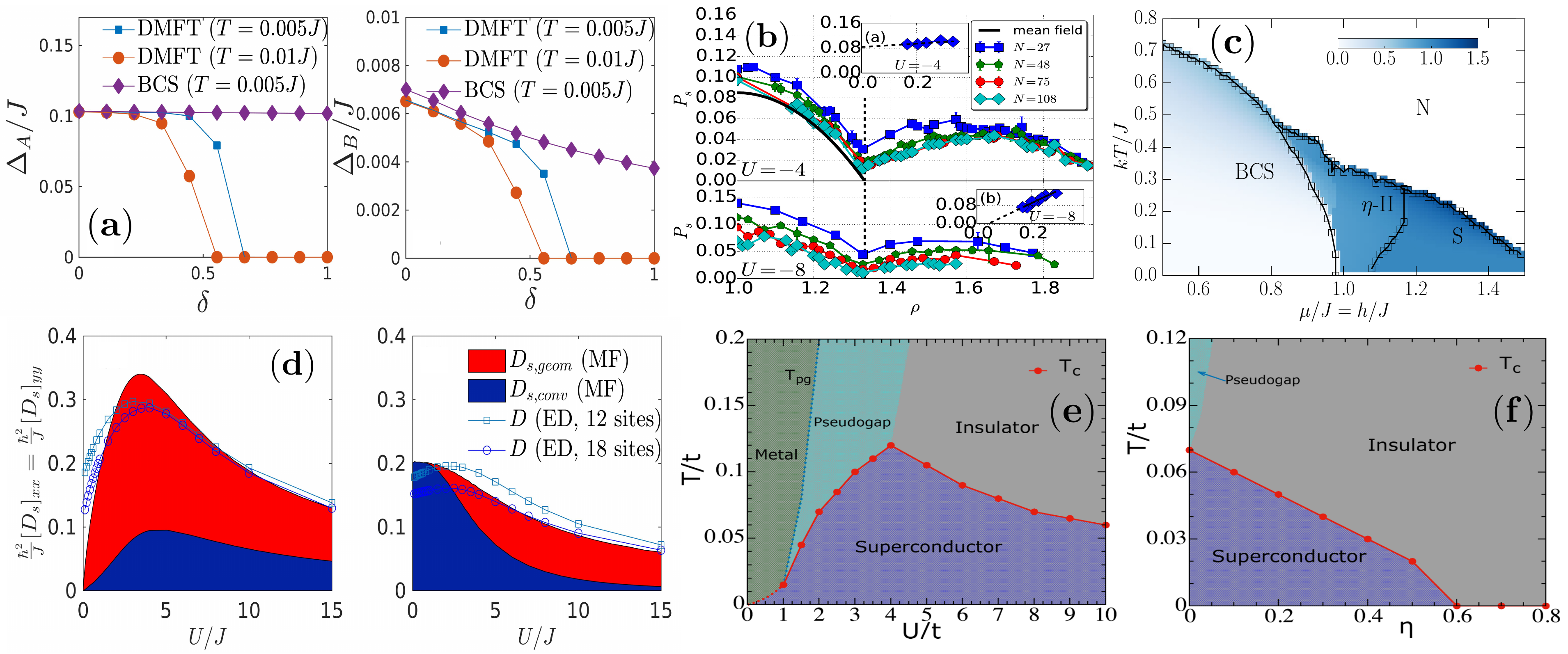}
\label{fig5}
\caption{Superconductivity in Lieb lattice. (a) Sub-lattice resolved superconducting order parameter $\Delta_{A}$ and $\Delta_{B}$ at different temperatures 
as a function of $\delta$, the staggered hopping (dimerization) parameter, as obtained using DMFT and mean field BCS theory, for a lattice filling of $\nu=1.5$ and interaction $U=-0.4J$ ($J=1$ is the reference energy scale). Figure adapted with permission from \cite{julku_prl2016}. Copyright 2016, American Physical Society. (b) Superconducting pair structure factor $P_{s}$ versus electron number density $\rho$ for $U=-4t$ and $U=-8t$, determined using DQMC. $P_{s}$ has a minimum at $\rho=4/3$ where the superconducting phase must compete with charge order, and is a maximum when $\rho=1$ and $\rho=5/3$. Inset: $P_{s}$ at $U=-4t$ and half filling, extrapolates to a non zero value in the thermodynamic limit. For $U=-8t$, the extrapolation is to zero value. Figure adapted with permission from \cite{scalettar_prb2014}. Copyright 2014, American Physical Society. (c) Phase diagram of the spin imbalanced attractive Hubbard model on the Lieb lattice, in the temperature-Zeeman field plane at a fixed interaction of $U=-4J$, as obtained using MFT. The phase diagram is mapped out along the flat band singularity line $\mu=h$, where $\mu$ is the chemical potential. Color intensity denotes the magnetic polarization and lines demarcate the phase boundaries. Figure adapted with permission from \cite{tylutki_prb2018}. Copyright 2018, American Physical Society. (d) Conventional $D_{s, conv}$ (blue) 
and geometric $D_{s, geom}$ (red) superfluid weight for a Lieb lattice at a filling of $\nu=1.5$ (left) and $\nu=2.5$ (right), as obtained using DMFT. The 
magnitude of the geometric weight is significantly larger than that of the conventional weight. Figure adapted with permission from \cite{julku_prl2016}. Copyright 2016, American Physical Society. (e) Thermal phase diagram showing the BCS-BEC crossover in the Lieb lattice, determined using SPA. The 
behavior of T$_{c}$ is nonmonotonic with peak at $U \sim 4t$ (where $t=1$ sets the reference energy scale). (f) Superconductor-insulator transition (SIT) across the strain ($\eta$)-temperature ($T$) plane in a Lieb lattice, at $U=2t$. The quantum phase transition occurs at a critical strain of $\eta_{c}=0.6$. The red (dotted) curve corresponds to the T$_{c}$ of the system. Figure adapted with permission from \cite{swain_prr2020}. Copyright 2020, American Physical Society.}
\end{center}
\end{figure*}

Superconductivity and the influence of flat electronic band on the same were investigated in the context of Kagome lattice as well, particularly in the 
recently discovered family of Kagome metals AV$_{3}$Sb$_{5}$ (where, A = K, Cs, Rb) \cite{ortiz_prm2019}. One of the pertinent issues is 
the pairing state symmetry of the Kagome superconductors which continues to be debated till date. An important work on the AV$_{3}$Sb$_{5}$ Kagome superconductors modelled in terms of a six-band tight binding Hamiltonian with multi-orbital density-density type interactions, Hund's coupling and based 
on the FRG theory,  suggested that over a range of interactions an $f$-wave triplet-pairing state is stabilized which gives way to a $d$-wave singlet pairing 
at stronger coupling, as shown in Figure 6(a). Moreover, a $p$-wave superconducting order is found to be sub-dominant across the regime of interactions \cite{wu_prl2021}.  In a similar spirit, a sub-lattice modulated superconductivity (SMS) viz. a pair density wave (PDW) superconducting order was shown to be stabilized in combination with a $d$-wave superconducting state, based on FRG calculations on a Kagome lattice. 
Such phases were claimed to be generic to lattices which can host sub-lattice modulated phases characterized by broken rotational symmetry rather than broken translational symmetry \cite{schwemmer_prb2024}. DQMC based estimate of superconducting pair structure factor $P_{s}$ on a Kagome lattice 
is shown in Figure 6(b). $P_{s}$ is non monotonic in terms of its dependence on the lattice filling $\rho$ and vanishes at $\rho=0, 4/3 and 2$ 
\cite{scalettar_prb2014}. The frustrated geometry of the Kagome lattice makes  unconventional pairing symmetry such as, $d$- or $f$-wave a likely choice, however based on DQMC an $s$-wave pairing state was found to be stabilized at low temperatures in a Kagome superconductor, with $T_{c} \sim 0.11t$ at $U=-4t$ and a chemical potential of $\mu=0.9t$, as shown in Figure 6(c) \cite{zhu_prr2023}.  

Considerable efforts have been invested to understand the Kagome superconductors based on the MFT \cite{yu_prb2012,yao_prb2025,lin_prb2024,tazai_pnas2024}. Using variational cluster approach (VCA) on $1/6$ hole doped Kagome lattice a chiral superconducting state $d_{x^{2}-y^{2}}+id_{xy}$ was found to be stabilized in the weak coupling ($U < 3t$) regime of the Hubbard model, followed by a disordered insulating state at intermediate coupling. The strong coupling regime ($U > 5.5t$) of this model was quantified by a chiral spin order \cite{yu_prb2012}. A self consistent MFT along with Ginzburg-Landau (GL) theory was used to study the on-bond attractive pairing on the Kagome lattice, exhibiting a stable 3${\bf Q}$ PDW phase, at the van Hove filling with chiral topological properties even in the absence 
of an explicit SOC (see Figure 6(d)) \cite{yao_prb2025}. A combination of GL theory and tensor network was utilized to understand the PDW order on a Kagome lattice exploring the vestigial ordered phases and topological defects \cite{lin_prb2025}. The inferences from this work were in relevance with the observed charge-$6e$ magnetoresistance oscillations in AV$_{3}$Sb$_{5}$ superconductors \cite{ge_prx2024}. A Boguliubov-de-Gennes MFT (BdG-MFT) analysis was carried out to understand the interplay between the superconducting and various CDW orders in Kagome materials. In particular, two dominant CDW configurations viz. trihexagonal and star-of-David patterns, involving charge bond order and chiral flux phase, with real and imaginary bond orders were explored in detail in presence of $s$-wave superconductivity \cite{lin_prb2024}. Moreover, on a multiorbital model,  GL theory based calculations were carried out to derive a magnetic field dependent free energy coupling the loop currents and bond orders \cite{tazai_pnas2024}. A beyond MFT paramagnon interference mechanism was used to understand the unconventional superconductivity in the Kagome metal AV$_{3}$Sb$_{5}$. Allowing for inter-sub-lattice scattering a spin fluctuations mediated smectic bond order is realized in this material with the fluctuating bond order serving as the required pairing glue for a $s$-wave pairing \cite{tazai_sciadv2022}.  
\begin{figure*}
\begin{center}
\includegraphics[height=6.5cm,width=10cm,angle=0]{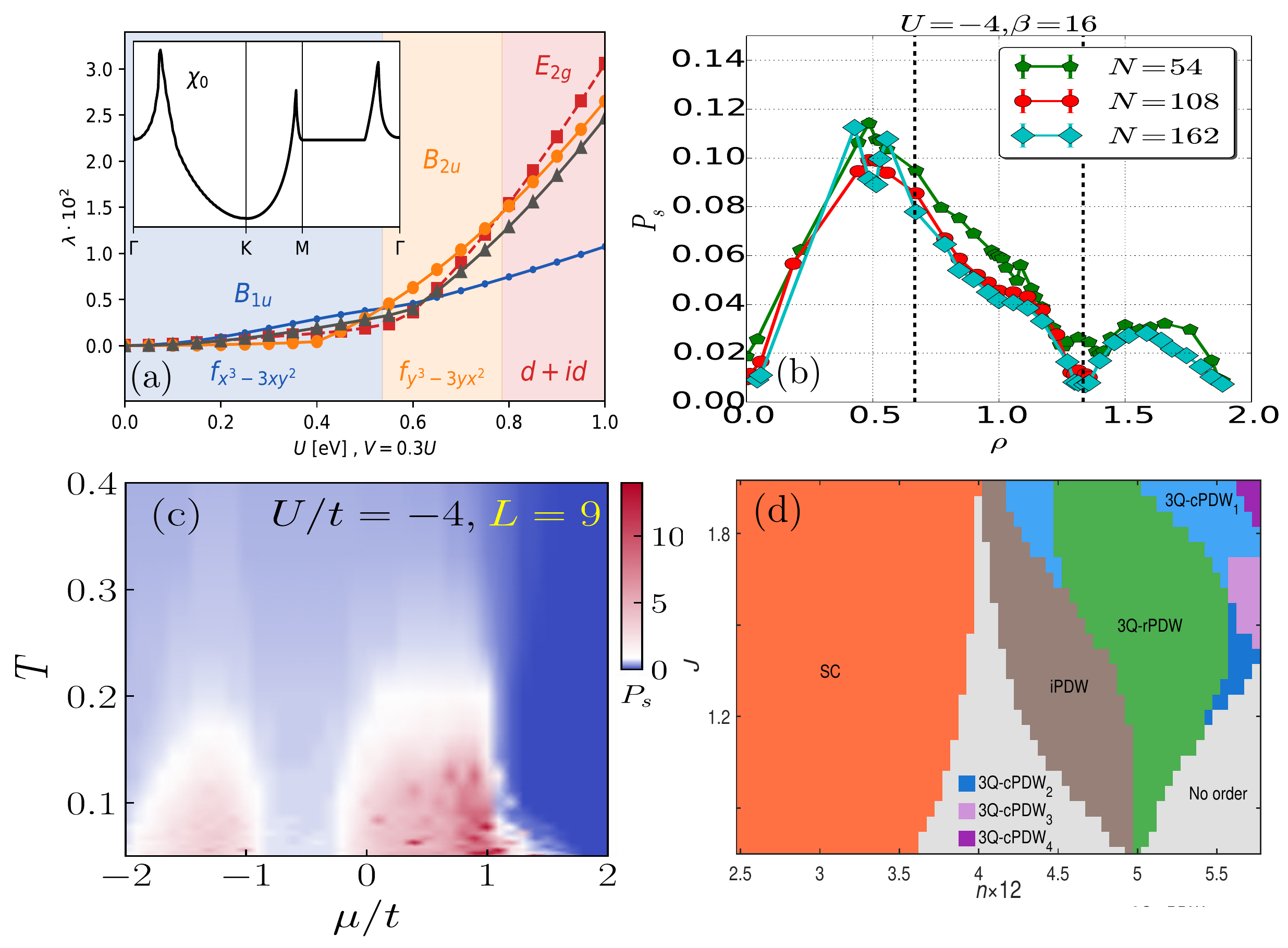}
\label{fig6}
\caption{Superconductivity in Kagome lattice: (a) Superconducting instability in the Kagome lattice as determined using FRG showing the pairing 
strength eigenvalues $\lambda$ for the dominant instabilities as a function of the interaction. Continuous (dashed) lines indicate triplet (singlet) pairing. 
Two distinct $f$-wave solutions dominate the weak interaction regime, $d$-wave solution dominates the larger $U$; $p$-wave solution is subleading 
but competitive at all interactions. Inset: largest eigenvalue trajectory of the bare susceptibility along the high symmetry path. Figure adapted with permission from \cite{wu_prl2021}. Copyright 2021, American Physical Society. Pairing structure factor $P_{s}$ versus the lattice filling $\rho$ at $U=-4t$ on the 
Kagome lattice, as obtained using DQMC. Figure adapted with permission from \cite{scalettar_prb2014}. Copyright 2014, American Physical Society. 
(c) The pair structure factor $P_{s}$ in the $T/t$-$\mu/t$ plane at $U=-4t$ for a Kagome lattice of size $L=9$. Figure adapted with permission from \cite{zhu_prr2023}. Copyright 2023, American Physical Society. (d) Mean field phase diagram in the plane of on-bond pairing interaction $J$ and lattice 
filling $n$ at $T=0.005$, on the Kagome lattice. The phases are demarcated in terms of colors, wherein SC corresponds to the uniform $d$-wave superconducting order and the various PDW phases indicate a multitude of pair density wave orders,  distinguished in terms of them being real and 
imaginary and their sub-lattice dependence \cite{yao_prb2025}. Figure adapted with permission from \cite{yao_prb2025}. Copyright 2025, American 
Physical Society.}
\end{center}
\end{figure*}

Apart from the ones discussed so far superconductivity on the Kagome lattice was studied in the weak coupling limit and close to the van Hove filling using analytic renormalization group (RG) approach and it was shown that sub-lattice interference suppresses the $d+id$ superconductivity while an increase in the T$_{c}$ can be achieved by including long range Hubbard interactions in the model \cite{kiesel_prb2012}. Non trivial response of the Kagome superconductor 
to non-magnetic disorder potential was reported based on symmetry analysis and model calculations. It was demonstrated that though spin-triplet states are fragile towards the disorder potential the spin-singlet states are only weakly pair breaking due to disorder even though the gap structure changes sign. The observations were attributed to the sub-lattice interference effect \cite{holbaek_prb2023}. The influence of correlated disorder on Kagome superconductors was studied based on the BdG theory wherein the effect of the sub-lattice geometry showed up as a bimodal distribution of the superconducting pairing field at strong disorder \cite{kiran_prb2024}. Combined experimental and theoretical studies carried out on the Kagome superconductor Lu(Ru$_{1-x}$Fe$_{x}$)$_{3}$Si$_{2}$ showed that magnetic impurities lead to strong suppression of the superconducting T$_{c}$ in this material resulting in emergent unconventional superconducting pairing with gap nodes \cite{mielke_prb2024}. 

The true pairing state of a Kagome superconductor is an open problem so far with the various theoretical approaches showcasing a rich phase landscape comprising of both conventional and unconventional paired phases such as, PDW, chiral phases, charge/valence bond entangled phases, loop currents 
and so on. At the same time the expanding literature on the experimental discovery and study of Kagome superconducting materials continue to pose new 
and higher challenges every day. 

\subsection{Magnetic order, Mott transition and thermal scales in flat band lattices}
Early works by Lieb showed that the large density of states at the Fermi level due to the flat band stabilizes a ferrimagnetic metal in the non interacting limit of the decorated square (Lieb) lattice, away from half filling \cite{lieb_prl1989}.  For a half filled lattice an infinitesimal repulsive interaction between the electrons open up a Mott gap at the Fermi level giving rise to a ferrimagnetic insulating state. This is in contrast to the Tasaki and Mielke lattices which hosts a ferromagnetic metal as the ground state even in presence of a finite repulsive interaction \cite{mielke_jphysmath1991,mielke2_jphysmath1991,tasaki_prl1992}. 
Using MFT and related techniques the magnetic phase diagram of the Lieb lattice as a function of the Hubbard repulsion $U$ was mapped out demarcating 
the ferrimagnetic, spiral, ferromagnetic and paramagnetic phases \cite{gouveia_jmmm2015}. Moreover, a stable ferromagnetic ground state was realized for $4/9$ filling of the Lieb lattice, based on the MFT analysis \cite{nie_pra2017}. Finite temperature phases were explored for the repulsive Hubbard model on the Lieb lattice using real space DMFT and continuous time QMC (CTQMC) techniques, bringing forth the NFL phases and the thermal scales associated with the Mott transition, as shown in Figure 7(a) \cite{kumar_prb2017}. Further, DQMC in conjunction with principle component analysis was used to investigate the metal-insulator transition in a $1/6$ filled Lieb lattice \cite{costa_prb2017} while the interplay of repulsive interaction, non magnetic disorder and particle number density on this lattice was shown to bring forth a metal-Anderson insulator transition, based on DQMC calculations \cite{li_prb2022}. 
The magnetic order and sub lattice magnetization were investigated both for a homogeneous and an inhomogeneous Lieb lattice using DQMC. The analysis was carried out in the context of cuprate with the corner and edge sites of the Lieb lattice corresponding to the $d$ and $p$-orbitals of CuO$_{2}$ planes of the cuprate, respectively, and the corresponding repulsive interactions being quantified as $U_{d}$ and $U_{p}$. The $U_{d} \neq U_{p}$ case simulates the Lieb geometry and it was shown based on DQMC study that at half filling a ferromagnetic insulator is realized for $U_{d}=0$ while for $U_{p}=0$ a non magnetic metal is stabilized \cite{costa_prb2016}. The corresponding ground state phase diagram exhibits a stable ferromagnetic phase as shown in Figure 7(b).

Recently it was suggested that the decorated square lattice geometry, such as the Lieb lattice is a natural selection to stabilize an altermagnetic ground state, particularly in presence of a mismatch in the onsite energy or chemical potential between the edge and the corner sites. Once again MFT unveiled exotic magnetic phases such as, vortex and block phases, quantified by specific spin structures,  at selected fillings of this altermagnetic setup as shown in 
Figure 7(c) \cite{kaushal_arxiv2024}. Based on SPA quantum Monte Carlo simulations it was observed that the symmetry protected bipartiteness of the 
Lieb lattice essentially promotes the altermagnetic correlations even in the absence of any bias such as, imbalanced site energy. Systematic annealing of the system at a fixed lattice filling spontaneously creates an imbalance in the fermionic densities between the edge and the corner sites,  leading to an emergent altermagnetic order \cite{karmakar_alm2025}. This observation of emergent altermagnetism opens up the possibility of its realization in engineered materials such as, MOFs and COFs, wherein the Lieb lattice geometry is a standard template. Altermagnets are materials which exhibits spin-split bands with vanishing global magnetization, reminiscent of ferro and antiferromagnetism, respectively \cite{smejkel_prx2022}. Non trivial spin-selective transport signatures are therefore expected from these materials and realizing the same in a MOF/COF setup will be a great step forward towards functional quantum materials. 

In comparison to the Lieb lattice the interacting Kagome lattice has been far more extensively studied owing to its relevance as a prototypical geometrically frustrated system which refuses to establish (quasi) long range order even at the lowest temperature \cite{balents_nature2010,savary_rpp2017,zhou_rmp2017}. 
This extensive research was however largely restricted to the strong coupling regime of Hubbard interaction corresponding to the insulating spin-$1/2$ quantum magnets.  The Kagome Hubbard Model (KHM) has served as the foundational framework for exploring the intricate interplay between the strong electronic correlations and geometric frustration in these materials, particularly in the strong-coupling Heisenberg limit where the precise nature of the ground state continues to remain a subject of debate, with the possible candidates ranging from Dirac spin liquid \cite{ran_prl2007,hermele_prb2008,ma_prl2008,he_prx2017,liao_prl2017,chen_scibul2018}, chiral spin liquid \cite{messio_prl2012,capponi_prb2019,bauer_natcom2014,he_prl2014,wietek_prb2015,gong_prb2015,messio_prl2017}, Z$_{2}$ spin liquid \cite{yan_science2011,depenbrock_prl2012,jiang_natphys2012,nishimoto_natcom2013,kolley_prb2015,mei_prb2017,lauchli_prb2019}, valence 
bond solid \cite{marston_jap1991,syromyatnikov_prb2002,nikolic_prb2003,singh_prb2007,budnik_prl2004,evenbly_prl2004,schwandt_prb2010,poiblanc_prb2010,poiblanc_prb2011}, to name a few. In contrast, the weak-coupling regime of the KHM has received little or no attention due to the absence of any suitable material realization in this regime. However, recent discoveries of Kagome metals—such as Mn$_3$Sn \cite{nakatsuji_nature2015,nayak_sciadv2016,kuroda_natmat2017,kimata_nature2019,collington_natcom2019,wuttke_prb2019}, Fe$_3$Sn$_2$ \cite{fenner_jpcm2009,hou_advmat2017,kang_nature2018,yin_nature2018,lin_prl2018,li_apl2019,li_prl2019,tanaka_prb2020}, Co$_3$Sn$_2$S$_2$ \cite{liu_natphys2018,wang_natcom2018,yin_natphys2019,liu_science2019,shen_apl2019,lachman_natcom2020}, Gd$_3$Ru$4$Al${12}$ \cite{nakamura_prb2018,matsumura_jpsj2019}, FeSn \cite{inoue_apl2019,kang_natmat2019,lin_prb2020,sales_prm2019}, and the AV$_3$Sb$_5$ family 
(A = K, Rb, Cs) \cite{ortiz_prm2019,ortiz_prl2020}—have reignited the interest in this under explored regime.  It was observed that for the interacting system 
the dispersion less electronic band brings forth novel strong correlation properties. Unlike the Lieb lattice the flat band in the non interacting Kagome lattice is 
a high energy band, away from the Fermi level and is therefore not expected to take part in dictating the electronic properties of these systems. However, numerical calculations based on SPA quantum Monte Carlo showed that electronic interaction leads to a systematic shift of the flat band towards the Fermi 
level, allowing it to participate in the electronic transport. Further, it was observed that the magnetic local moments in the flat band tends to localize the itinerant fermions at and close to the Fermi level, leading to an Anderson insulator like state sans any potential disorder, termed as the flat band induced insulator 
(FI), as shown in Figure 7(f) \cite{shashi_kagome2025}. The observations are found to be of significant relevance to the NFL signatures observed in the electronic transport measurements on Ni$_{3}$In, a Kagome metal, exhibiting Kondo like coupling between the local moments originating from the flat band 
and the itinerant fermions at the Fermi level \cite{checkelsky_natphys2024}.  A growing body of work employing techniques such as DMFT and its extensions  \cite{tsunetsugu_jpcm2007,asano_prb2016,hatsugai_prl2019,ohashi_prl2006}, DQMC \cite{maekawa_prl2005,janson_prb2021,paiva_prb2023}, D$\Gamma$A \cite{held_rmp2018,held_prb2021,toschi_prb2016}, and VCA  \cite{asano_prb2016} has established a general picture of the low-temperature behavior of the KHM at half filling. These studies indicate that the model remains metallic at weak and intermediate interaction strengths, undergoing a first-order metal-insulator transition (MIT) at a critical interaction strength $U_c \sim 5t$–$11t$, with the precise value depending on the computational method and the treatment of the correlation effects. Figure 7(d) shows the estimate of T$_{c}$ as obtained based on the interaction dependence of the quasiparticle weight determined 
using DQMC and DMFT \cite{janson_prb2021}.  

Due to the strong correlations and inherent magnetic frustration, accurate characterization of the Kagome materials require non-perturbative numerical 
methods, though frustration significantly limits the applicability of most of the conventional techniques, especially at low temperatures. On the other hand, 
it is the low temperature regime which is of utmost importance, wherein competing low-energy excitations play a dominant role in promoting a stable ground state out of the highly degenerate energy manifold. To probe the ground state, specialized approaches like FRG \cite{thomale_prr2024} and DMRG \cite{zhu_prbl2021} have been employed, focusing on the zero-temperature limit. Recent DMRG results suggest the presence of two distinct critical points in 
the half filled KHM: a translation symmetry-breaking insulating phase near $U \sim 5.4t$, and a quantum spin liquid phase at $U \sim 7.9t$, see Figure 7(e) \cite{zhu_prbl2021}. Despite these advances, our understanding of the low but finite temperature behavior of the half-filled KHM remains limited. Current insights are largely based on extrapolations from either high-temperature or $T = 0$ results, yet it is in this intermediate regime—dominated by short-range correlations—that novel phase crossovers may emerge. This temperature window is also of particular experimental relevance, as it exhibits spectroscopic, and transport behavior that deviates markedly from the conventional Fermi liquid theory. Recently, SPA quantum Monte Carlo was applied to systematically investigate the 
low temperature regime of the half-filled KHM across the complete range of interactions, for the first time. Analysis of the thermodynamic, spectroscopic, and transport data revealed that in the weak-coupling regime ($0 < U \le U_{c1} \sim 3.6t$) itinerant electrons become transiently localized due to interactions with thermal bosonic fluctuations of flat-band-induced local moments. This leads to suppressed charge transport and the emergence of a NFL flat-band insulating  state. At intermediate-couplings, an NFL metallic phase appears, with $U_{CM} \sim 4.0t$ marking the onset of antiferromagnetic correlations, followed by a 
first-order MIT to an antiferromagnetic Mott insulator (AF-MI), at $U_{c2} \sim 4.4t$. These findings were supported by transport and optical data, where features such as deviation from the $\rho_{xx} \propto T^2$ scaling and a displaced Drude peak (DDP) in the optical conductivity are consistent with the proposed dynamic localization scenario \cite{shashi_kagome2025}. The inferences made based on the numerical calculations found credence via the electrical and 
optical transport measurements on Ni$_3$In \cite{checkelsky_natphys2024} and CsV$_3$Sb$_5$ \cite{tsirlin_prb2021}, exhibiting the salient features of NFL physics. 
\begin{figure*}
\begin{center}
\includegraphics[height=6.5cm,width=14cm,angle=0]{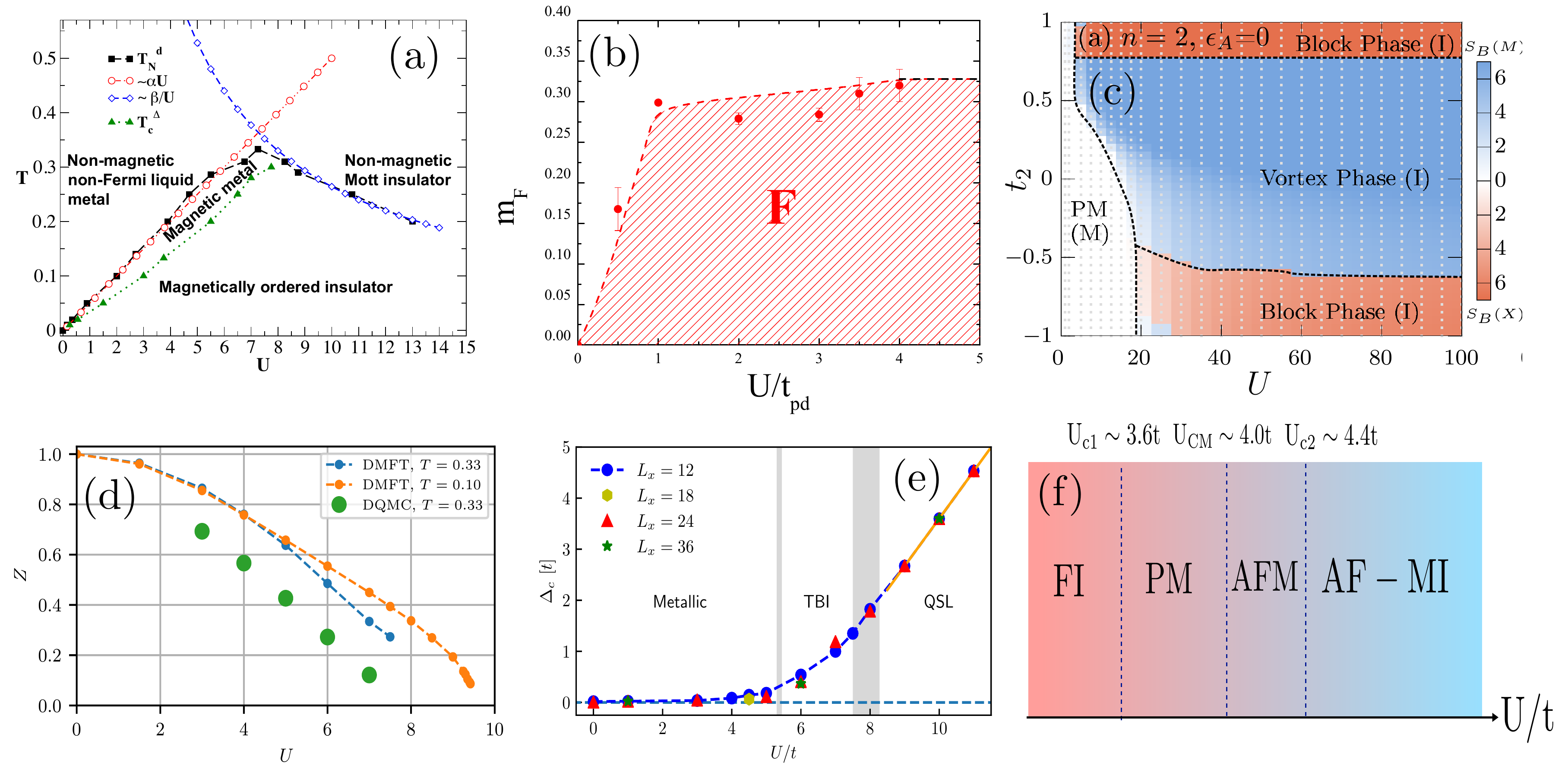}
\label{fig7}
\caption{Magnetic phases and Mott transition in flat band lattices. (a) Finite temperature phase diagram for the repulsive Hubbard model on the Lieb lattice, 
as obtained using DMFT. Various phases are marked individually and the curves indicate the thermal transition and crossover scales \cite{kumar_prb2017}. 
Figure adapted with permission from \cite{kumar_prb2017}. Copyright 2017, American Physical Society. (b) Global ferromagnetic order parameter as a function of the on-site repulsion $U$, obtained using DQMC for the repulsive Hubbard model on the Lieb lattice. Figure adapted with permission from \cite{costa_prb2016}. Copyright 2016, American Physical Society. (c) Phase diagram showing altermagnetic phases on a modified Lieb lattice. $t_{2}$ corresponds to the next nearest neighbor hopping on a Lieb lattice and $U$ is the repulsive Hubbard interaction \cite{kaushal_arxiv2024}. Figure adapted from \cite{kaushal_arxiv2024}. (d) Quasiparticle renormalization weight ($Z$) quantifying the metal-insulator transition in the half filled Kagome Hubbard model, as determined using DMFT and DQMC \cite{janson_prb2021}. Figure adapted with permission from \cite{janson_prb2021}. Copyright 2021, American Physical Society. (e) Charge gap ($\Delta_{c}$) as a function of $U/t$, showing MIT in the half filled Kagome Hubbard model, as obtained using DMRG \cite{zhu_prbl2021}. Figure adapted with permission from \cite{zhu_prbl2021}. Copyright 2021, American Physical Society. (f) Low temperature phase 
diagram ($T=0.01t$) of the half filled Kagome Hubbard model, mapped out based on SPA. The various phases include: flat band induced localized insulator (FI), paramagnetic metal (PM), antiferromagnetic metal (AFM) and antiferromagnetic Mott insulator (AF-MI) \cite{shashi_kagome2025}. Figure adapted from \cite{shashi_kagome2025}.}
\end{center}
\end{figure*}

This work brought forth the existence of the low temperature ($T \neq 0$) transiently localized FI phase in the weak coupling regime of the half filled KHM,  contrary to the existing consensus of a metal. Although the gapless nature of this phase might initially imply metallic behavior, the transport signatures clearly indicates insulating tendencies. A gapless insulating phase is a natural to disordered quantum systems, in the absence of an explicit external disorder potential however, similar localization tendencies can stem from the intrinsic disorder of the randomly fluctuating thermal bosonic fields \cite{ciuchi_scipost2021,fratini_prb2023,ciuchi_arxiv2024}. The analysis relies on the adiabatic approximation, where the bosonic field evolves slowly enough that the fermions perceive it as a static, spatially disordered background—effectively classical in nature. The resulting fermionic localization is transient, persisting over short timescales, above the coherence temperature $T_{FL}$, and is expected to manifest at frequencies exceeding the characteristic bosonic frequency of the system \cite{ciuchi_scipost2021,fratini_prb2023,ciuchi_arxiv2024}. More recently, the inter-convertibility of the line graph lattices with Lieb and Kagome geometries was explored in significant detail with the prospect of straintronics. While the emergent topological characteristics of such an interconversion are well known \cite{jiang_prb2019,lang_pra2023}, this work demonstrated for the first time the strain controlled realization of exotic quantum correlated phases such as, flat band induced transiently localized insulator, (alter)magnetic insulator and NFL metal, across the Lieb-Kagome interconversion \cite{shashi_kagome_lieb}.  Further, the transport properties showed strain dependent variable resistivity and optical conductivity scaling exponents, salient not just to a NFL metal but to a dynamically localized phase, as well \cite{shashi_kagome_lieb}. 

\subsection{Quantum MOFs: strongly correlated MOFs with flat band}
The majority of the functionalities and properties of the MOFs, COFs and related classes of materials discussed in this review are based entirely on their 
lattice topology and on the choice of the functional building blocks, leaving the ``quantumness'' of these materials largely unexplored. On the other hand the prototypical MOFs and COFs (with their Lieb and/or Kagome lattice geometry) comprise of flat electronic band(s), making them an ideal premise for the 
interplay of strong electronic correlations and quenched energy dispersions. It is only very recent that the focus of the material science research has begin 
to shift towards the electronic degrees of freedom in these materials, giving rise to what is now known as {\it quantum MOFs/COFs} \cite{huang_smallscience2024}. Recently, locally confined magnetic moments arising out of the electron-electron Coulomb interaction in a MOF comprising of 
di-cyano-anthracene (DCA) molecules arranged in a Kagome geometry via coordination with copper (Cu) atoms on a silver surface [Ag(111)], 
was reported and probed using the Kondo effect \cite{kumar_advfuncmat2021}. In a similar spirit, electronic correlation induced magnetism was reported in the 2D Kagome MOF, 9, 10-dicyanoanthracene-copper (DCA-Cu) on Ag(111), Cu(111) and graphite substrate etc., based on density functional theory and mean field based calculations \cite{field_npjcompmat2022}. In particular, the effect of the choice of the substrate and the response of the strong electronic correlation in the MOF to the tuneable external perturbations such as, an applied strain and an applied electric field were probed, bringing forth correlation dependent thermodynamic phase transitions arising out of the interplay between the geometric frustration, electron filling and external perturbations in these systems. 

The plausibility of Mott MIT, a prototypical strong correlation phenomena, was observed on single layer DCA$_{3}$Cu$_{2}$, a Kagome MOF 
\cite{lowe_natcom2024}. STM and spectroscopy revealed the Mott gap of $~200$meV in this MOF, the MIT was achieved via the combination of template induced and STM probe-induced local electrostatic gating, which locally altered the electron population of the Kagome bands of the MOF.  Further, an insulator-metal transition was realized in MCl$_{2}$(pyrazine)$_{2}$ coordination solid with Lieb lattice structure, via the replacement of the metal ions. While VCl$_{2}$(pyz)$_{2}$ with divalent V$^{II}$ ions is an antiferromagnetic insulator for $T \lesssim 120K$, TiCl$_{2}$(pyz)$_{2}$ with trivalent Ti$^{III}$ ions exhibit a large electronic conductivity of $\sim$ 5.3Scm$^{-1}$ at room temperature, as well as a large positive magnetoresistance at low temperatures \cite{perlepe_natcom2022}. Moreover, the temperature dependence of the resistivity suggests that the underlying state of this material is a NFL with pronounced deviation from the $\rho_{xx} \propto T^{2}$ dependence. Unconventional superconductivity mediated by the geometric frustration induced spin fluctuations was reported in $\pi-d$ conjugated benzenehexathiol based 2D MOF [Cu$_{3}$(C$_{6}$S$_{6}$)]$_{n}$ (Cu-BHT) with Kagome geometry \cite{takenaka_sciadv2021}. Based on electrical transport and reflectivity measurements it was established that unlike the previous DFT based predictions of a $s$-wave superconductivity \cite{zhang_nanolett2017}, Cu-BHT is a strongly correlated superconductor proximate to a quantum spin liquid state and has a transition temperature of  T$_{c} \sim $ 0.25K. The observation was backed up by the anomalous behavior of the normal state heat capacity and magnetic susceptibility in this material, attesting to an underlying NFL metal \cite{takenaka_sciadv2021}. Quantum MOFs/COFs are in their inceptive stage and holds immense potential for the material engineering and applications. It was recently proposed that MOFs serve as suitable candidate to exhibit quantum buckling, wherein under strain, deformation or buckling of the structure can take place in the nanoscale, a property relevant in the context of mechanical qubits \cite{huang_smallscience2024,geilhufe_nanolett2021}.  Quantum buckling of a MOF comprising of Zn$_{4}$O clusters and BDC molecules was modelled using a transverse field Ising Hamiltonian, 
\begin{eqnarray}
H &= & -t\sum_{i}\sigma_{i}^{x} - \sum_{ij}J_{ij}\sigma_{i}^{z}\sigma_{j}^{z}
\end{eqnarray}
where, $J_{ij}$ is an exchange energy between molecules $i$, $j$ and $\sigma^{x}$, $\sigma^{y}$, $\sigma^{z}$ are the Pauli matrices. Mean field analysis of this Hamiltonian with $J_{ij}=J$ being the nearest-neighbor interaction reveals a phase diagram comprising of three phases as: normal, parabuckling and ferrobuckling, determined 
by the barrier height, exchange energy $J$, tunnelling strength $t$ and molecule's ground state energy. The quantum states emerge when the barrier height exceeds the ground state energy, on the other hand, an ordered buckling is prevented due to quantum fluctuations if the tunnelling dominates. Otherwise, a collective buckling can take place with a transition temperature of $T_{c} \approx 4J$ \cite{huang_smallscience2024,geilhufe_nanolett2021}. 

\section{Simulating the quantum correlations in flat bands: beyond mean field theory}
In this review we have presented the various aspects of materials comprising of dispersion less electronic bands. Starting from the mathematical construct 
of the non interacting lattice structure of these materials we have discussed the impact of strong electronic correlations both from the 
experimental and theoretical point of view. It was argued that in order to capture the physics of many body correlations in these materials a non perturbative 
numerical approach is required, however most of the existing approaches are limited in their applications owing to various computational bottlenecks. Further, both in the context of unconventional superconductivity and Mott insulating properties of flat band materials it was suggested that the SPA quantum Monte 
Carlo technique can prove to be a suitable choice, a judicial approximation which while alleviating many of the existing computational bottlenecks, can capture the low and high temperature properties of these materials with reasonable accuracy \cite{evenson_jap1970,dubi_nature2007,karmakar_pra2016,karmakar_prb2022,swain_prr2020,dagotto_prl2005}. 

In this section we outline the numerical framework for the SPA quantum Monte Carlo technique while using the 2D repulsive Hubbard Hamiltonian as the prototypical strong correlation model. The framework can however be easily generalized to address other many body Hamiltonians. The prototypical 2D 
repulsive Hubbard model on a Kagome lattice is defined as, 
\begin{eqnarray}
\hat{H} = -t \sum_{\langle ij  \rangle,\sigma}(\hat c^{\dagger}_{i,\sigma}\hat c_{j,\sigma} + h.c) - \mu \sum_{i,\sigma} \hat n_{i\sigma} \nonumber + U \sum_{i}\hat n_{i,\uparrow}\hat n_{i,\downarrow} \nonumber \\ 
\end{eqnarray}
Here, $t_{ij}=t$ represents the nearest-neighbor hopping on a Kagome lattice, with $t=1$ setting the reference energy scale of the problem. The parameter $U > 0$ denotes the on-site repulsive Hubbard interaction.  The chemical potential $\mu$ is adjusted to maintain the half filled lattice. To make the model numerically tractable,  Hubbard-Stratonovich (HS) transformation is applied to decompose the interaction term \cite{hs1,hs2}. This introduces two bosonic auxiliary fields: a vector field ${\bf m}_{i}(\tau)$ and a scalar field $\phi_{i}(\tau)$, which couple to the spin and charge densities, respectively. The introduction of these auxiliary fields preserves spin-rotation invariance, retains the Goldstone modes, and allows one to capture the Hartree-Fock theory at the saddle point. In terms of the Grassmann fields $\psi_{i\sigma}(\tau)$, we rewrite the interaction as follows:
\begin{eqnarray}
\exp[U\sum_{i}\bar\psi_{i\uparrow}\psi_{i\uparrow}\bar\psi_{i\downarrow}\psi_{i\downarrow}] & = & \int {\bf \Pi}_{i}
\frac{d\phi_{i}d{\bf m}_{i}}{4\pi^{2}U}{\exp}[\frac{\phi_{i}^{2}}{U}+i\phi_{i}\rho_{i}+\frac{m_{i}^{2}}{U} -2{\bf m}_{i}.{\bf s}_{i}]
\end{eqnarray}

where, the charge and spin densities are defined as, $\rho_{i} = \sum_{\sigma}\bar\psi_{i\sigma}\psi_{i\sigma}$ and ${\bf s}_{i}=(1/2)\sum_{a,b}\bar \psi_{ia}{\bf \sigma}_{ab}\psi_{ib}$, respectively. Upon applying the HS transformation, the partition function of the system takes a form that 
facilitates the integration over the fermionic degrees of freedom. The partition function is given by:

\begin{eqnarray}
{\cal Z} & = & \int {\bf \Pi}_{i}\frac{d\bar\psi_{i\sigma}d\psi_{i\sigma}d\phi_{i}d{\bf m}_{i}}{4\pi^{2}U}
\exp[-\int_{0}^{\beta}{\cal L}(\tau)d\tau]
\end{eqnarray}
where the Lagrangian density, ${\cal L}(\tau)$, is defined as:
\begin{eqnarray}
{\cal L}(\tau) & = & \sum_{i\sigma}\bar\psi_{i\sigma}(\tau)\partial_{\tau}\psi_{i\sigma}(\tau) + H_{0}(\tau) 
 +\sum_{i}[\frac{\phi_{i}(\tau)^{2}}{U}+(i\phi_{i}(\tau)-\mu)\rho_{i}(\tau) \nonumber \\ && +
\frac{m_{i}(\tau)^{2}}{U} -2{\bf m}_{i}(\tau).{\bf s}_{i}(\tau)]
\end{eqnarray}
Here $H_{0}(\tau)$ represents the kinetic energy contribution.
 
The integral over $\psi$ is now quadratic, but it introduces the need for an additional integration over the fields  ${\bf m}_{i}(\tau)$ and $\phi_{i}(\tau)$. The 
weight factor for the ${\bf m}_{i}$ and $\phi_{i}$ configurations can be determined by integrating out the fermionic fields $\psi$ and $\bar \psi$. Once these weighted configurations are determined, one can proceed to compute the fermionic properties. The partition function, following the procedure outlined above, 
is given by:
{\begin{eqnarray}
{\cal Z} & = & \int {\cal D}{\bf m}{\cal D}{\phi}e^{-S_{eff}\{{\bf m},\phi\}}
\end{eqnarray}}
The effective action $S_{eff}\{{\bf m}, \phi\}$ reads as:
\begin{eqnarray}
S_{eff}\{{\bf m}, \phi\} & = & \log Det[{\cal G}^{-1}\{{\bf m},\phi\}] + \frac{\phi_{i}^{2}}{U} +
\frac{m_{i}^{2}}{U}
\end{eqnarray}
where ${\cal G}$ represents the electron Green's function in the background of $\{{\bf m}_{i}, \phi_{i}\}$. The weight factor for an arbitrary space-time configuration $\{{\bf m}_{i}(\tau), \phi_{i}(\tau)\}$ involves the computation of the fermionic determinant in that specific background. If the auxiliary fields 
are expressed in terms of their Matsubara modes, ${\bf m}_{i}(\Omega_{n})$ and $\phi_{i}(\Omega_{n})$, the various approximations become apparent 
and can be directly compared:
 
\begin{itemize}
\item[1.]{{\bf Determinant Quantum Monte Carlo (DQMC)} preserves the full dependence of ${\bf m}$ and $\phi$ on both ``i'' and $\Omega_n$, and computes 
the logarithm of the determinant of ${\cal G}^{-1}\{{\bf m}, \phi\}$ iteratively for importance sampling. This method is applicable at all temperatures but does not easily provide real-frequency properties. Moreover, in many quantum materials, DQMC faces challenges in the low-temperature regime due to the fermionic sign problem. This issue is especially pronounced in magnetically frustrated systems, such as the Kagome lattice. Additionally, the high computational cost of DQMC limits its use to small system sizes, which results in significant finite-size effects in the outcomes.}
\item[2.]{{\bf Homogeneous Mean-Field Theory} is time-independent and completely neglects fluctuations. It approximates the auxiliary fields by their mean values and minimizes the free energy, i.e., ${\bf m}_i(\Omega_n) \rightarrow |m|$ and $\phi_i(\Omega_n) \rightarrow |\phi|$. In contrast, inhomogeneous Hartree-Fock Mean-Field Theory accounts for spatial fluctuations in the magnitudes of ${\bf m}_i$ and $\phi_i$, but ignores angular fluctuations, i.e., ${\bf m}_i(\Omega_n) \rightarrow |m_i|$ and $\phi_i(\Omega_n) \rightarrow \phi_i$. For nonzero temperatures ($T \neq 0$), this approximation breaks down beyond the weak-coupling regime.}
\item[3.]{{\bf Static Path Approximation (SPA)} retains the full spatial dependence of ${\bf m}$ and $\phi$, but only considers the $\Omega_n = 0$ mode, i.e., ${\bf m}_i(\Omega_n) \rightarrow {\bf m}_i$ and $\phi_i(\Omega_n) \rightarrow \phi_i$. This approximation includes classical fluctuations of any magnitude but ignores quantum fluctuations ($\Omega_n \neq 0$). The different temperature regimes are as follows: (a) At $T = 0$, since classical fluctuations vanish, SPA reduces to standard Hartree-Fock mean-field theory, (b) At $T \neq 0$, the approach considers  not just the saddle-point configuration but all configurations weighted by $e^{-H_{\text{eff}}}$ (Eqn.\ref{eqn}), which leads to faster suppression of order compared to mean-field theory, (c) At high $T$, since the $\Omega_n = 0$ mode dominates the partition function, the SPA becomes exact as $T \to \infty$.}
\item[4.]{{\bf Dynamical Mean-Field Theory (DMFT)} retains the full dynamics but approximates ${\bf m}$ and $\phi$ at a single site, i.e., ${\bf m}_i(\Omega_n) \rightarrow {\bf m}(\Omega_n)$ and $\phi_i(\Omega_n) \rightarrow \phi(\Omega_n)$. This approximation is exact in the limit of infinite spatial dimensionality ($D \rightarrow \infty$), where $D$ represents the spatial dimension.}
\end{itemize}

Following the Static Path Approximation (SPA), the field $\phi_i(\tau)$ is frozen to its saddle-point value, $\phi_i(\tau) = \langle n_i \rangle U / 2$, where $\langle n_i \rangle$ represents the fermionic number density. This leads to a model where fast-moving fermions interact with a slow, spatially fluctuating random background of the classical field ${\bf m}_i$. With these approximations, the effective Hamiltonian becomes a coupled spin-fermion model, given by: 
\begin{eqnarray}
H_{eff} & = & -t\sum_{\langle ij\rangle, \sigma}[c_{i\sigma}^{\dagger}c_{j\sigma}+h.c.] 
-\tilde \mu \sum_{i\sigma} \hat n_{i\sigma}  - \frac{U}{2}\sum_{i}{\bf m}_{i}.{\bf \sigma}_{i} + \frac{U}{4}\sum_{i}m_{i}^{2}
\label{eqn}
\end{eqnarray}
Here, $\tilde{\mu} = \sum_i (\mu - \langle n_i \rangle U / 2)$, and the final term in the Hamiltonian accounts for the stiffness cost of the now classical field 
${\bf m}_{i}$. Note that ${\bf \sigma}_{i} = \sum_{a,b} c_{ia}^{\dagger} \sigma_{ab} c_{ib}= {\bf s}_i$ represents the spin operator.

The random background configurations of ${{\bf m}_i}$ are generated numerically through Monte Carlo simulations, following the Boltzmann distribution:
\begin{eqnarray}
P\{{\bf m}_{i}\} \propto Tr_{c,c^{\dagger}}e^{-\beta H_{eff}}
\end{eqnarray}
For large and random configurations, the trace is computed numerically by diagonalizing $H_{\text{eff}}$ for each attempted update of ${\bf m}_i$. 
The equilibrium configurations are obtained using the Metropolis algorithm, which are then used to compute the various fermionic correlation functions. 
 
\section{Conclusions and Outlook}
This review explores the physics of quantum correlated systems with flat electronic bands, focusing on the Lieb and Kagome lattices—two prototypical 
examples distinguished by their unique topological and geometrical characteristics. In such systems, the suppression of kinetic energy inherent to flat 
bands enhances the impact of electron–electron interactions, creating a fertile ground for the emergence of strongly correlated quantum phases. 
Consequently, flat band lattices provide an ideal platform for both theoretical and experimental investigations into unconventional states of matter, 
including magnetism, superconductivity, and Mott insulating behavior.

We present a broad survey of various engineered platforms where flat electronic bands can be realized. These include ultracold atoms in optical lattices, 
where precise control over lattice geometry and interaction strength enables simulation of idealized tight-binding models; photonic waveguides and 
metamaterials, where the band structure can be tuned to emulate electronic flat bands; scanning tunnelling microscopy (STM)-based manipulation of 
atomic-scale electronic lattices; and low-dimensional coordination polymers, particularly metal-organic frameworks (MOFs) and covalent-organic 
frameworks (COFs), which offer chemically tuneable platforms with intrinsic flat bands.

The theoretical underpinnings of flat band formation are discussed in terms of tight-binding models, with a particular emphasis on the lattice geometry, 
orbital character, and symmetry conditions required to generate dispersionless bands. We outline the role of localized Wannier states and destructive 
interference mechanisms that lead to the formation of compact localized states, a hallmark of flat band systems. Beyond the single-particle picture, 
we systematically examine the many-body physics that emerges when interactions are introduced into flat band systems. Strong correlations in such 
systems can give rise to a rich landscape of quantum phases, including unconventional superconductivity, Wigner crystallization, and flat-band 
ferromagnetism. We focus on two key correlated phases—superconductors and Mott insulators—and analyze their stability and characteristics in 
flat band settings.

Our discussion emphasizes the Lieb and Kagome lattices, which are not only theoretically appealing but have also been realized or are within reach 
in several experimental platforms. These lattices exhibit nontrivial topology, frustration, and flat band features that make them especially relevant for 
the study of emergent quantum phenomena. Finally, we highlight a non-perturbative numerical framework—viz SPA quantum Monte Carlo technique—that enables quantitative analysis of interaction-driven phases and phase transitions in flat band systems. Such an approach is crucial for exploring regimes beyond 
the reach of conventional perturbative techniques and for making direct contact with experimental observations. We discussed how SPA circumvents 
some of the computational bottlenecks of the existing numerical approaches and unveil exotic quantum phases in these materials, such as, flat band 
induced transiently localized gapless insulator. 

Flat band materials are a relatively new area in condensed matter physics, and many fundamental questions about their behavior under various physical conditions remain open. These include the potential application of the flat band systems as functional materials, interplay between disorder and interaction 
effects, collective modes and light-matter interaction, to name a few.  Addressing these open problems requires both theoretical innovation and advanced experimental techniques. We touch upon some of these issues in this section. 

\subsection{Magnetoelectric coupling and flat band multiferroics}
There has been growing interest and technological demand in exploring the quantum materials for their prospective functionalities and multifaceted applications, 
in response to the external perturbations as well as electronic interaction. For example, based on continuum model Hamiltonian calculations it was recently proposed that twisted transition metal dichalcogenides (TTMD)s hosts multiferroic order as a function of electronic interaction - ferroelectricity arising out of 
layer polarization and ferromagnetism arising due to spin-valley polarization, as shown in Figure 8(a) \cite{abouelkomsan_prl2024}. There have been 
several proposals based on the DFT and related calculations harnessing the intrinsic magnetoelectric coupling of natural and artificial 2D van der Waals materials. MXene Hf$_{2}$VC$_{2}$F$_{2}$ monolayer was identified as a type-II multiferroic wherein ferroelectricity originates from the magnetic coupling \cite{zhang_jamchemsoc2018}, similarly in the context of 2D multiferroic ReWCl$_{6}$ it was demonstrated that the magnetic transition between the ferro 
and antiferromagnetic order can be realized via the applied electric field controlled flipping of polarization \cite{xu_prb2020}. Further, it was suggested that in CrPSe$_{3}$ the ferro and antiferroelectric phases can host topological magnetic vortices such as, meron pairs \cite{gao_prml2022}. Using a combination of experiment, model calculations and simulations chirality controlled electrical polarization was brought forth in NiI$_{2}$ \cite{song_nature2022}. In a similar 
spirit, dynamical magnetoelectric coupling was observed in NiI$_{2}$ in the form of giant terahertz magnetoelectric oscillations of chiral domains \cite{gao_nature2024}. Moreover, multiferroic tunnel junctions were formed using 2D ferromagnetic Fe$_{n}$GeTe$_{2}$ ($n = 3, 4, 5$) electrodes and 2D ferroelectric In$_{2}$Se$_{3}$ barrier layers \cite{su_nanolett2020}. More recently, flat band induced intrinsic magnetoelectric coupling was observed in 
Kagome van der Waals heterostructure CrGeTe$_{3}$/Nb$_{3}$Cl$_{8}$, based on first principle calculations \cite{lu_prb2025}.   

Though incipient and relatively unexplored the prospects of harnessing the magnetoelectric properties in flat band materials is immense, particularly since these materials owing to their lattice geometry are potential hosts of altermagnetic correlations. For example, magnetoelectric properties arising out of lattice-fermion coupling, such as magneto-piezoelectric effect (MPE) and superconducting piezoelectric effect (SCPE) if realized in a flat band set up with altermagnetic 
correlations instead of the standard approach of SOC,  should provide a non relativistic avenue for spin selective electronic transport \cite{chazono_prb2022}. 
In a similar spirit, another prospective functional application of these materials as a multiferroic is in the form of ferroelectric superconductors (see Figure 8(b)) and ferrolectric Mott insulator wherein the polar lattice distortion of the system could be modelled in terms of the emergent altermagnetic coupling in flat band materials, rather than via a SOC \cite{kanasugi_prb20218}. Yet another avenue to leverage the non trivial band structure of the flat band materials is by understanding the linear and non linear electronic transport properties of these systems. The important question that one needs to ask in this context is the role of geometric weight and quantum metric in the linear such as, anomalous Hall effect (AHE) \cite{nagaosa_rmp2010}, spin Hall effect (SHE) \cite{sinova_rmp2015}, large magnetoresistance \cite{manchon_rmp2019} and non linear properties such as, nonlinear Hall effect (NHE) \cite{kang_natmat2019}, Edelstein/Rashba effect \cite{manchon_rmp2019}, inverse Edelstein or spin galvanic effect \cite{shen_prl2014}, superconducting diode effect \cite{daido_prl2022} etc, induced by the current and temperature gradients. From a broader perspective, rationalizing the transport and magnetoelectric properties in flat band materials, in terms of their microscopic origin will open up the prospect of their application as altermagnetic systems, which can be designed and customized in engineered material platforms.  
 
\subsection{Interplay of disorder and correlation}
Anderson's theorem says that for a non-interacting system in 1D and 2D even an infinitesimal amount of non magnetic potential disorder localizes the energy states of the particles, while a 3D system requires a critical disorder strength to undergo metal-insulator transition \cite{anderson_disorder}. This description however breaks down in presence of electronic interaction in the system wherein the competition between the delocalizing tendency of the electronic interaction and localizing effect of potential disorder dictates the nature of the underlying phase, often bringing forth novel phases such as, correlated NFL metal \cite{karmakar_prb2022}. Itinerant electrons in flat bands are effectively strongly correlated, as has been already discussed, while at the same time the absence 
of dispersion in such bands lead to strong localization effects resulting in CLS states \cite{flack_advphysx2018}. The crucial question that stems from this seemingly contrasting scenarios is what is the nature of the quantum phase stabilized via the interplay of electronic interaction and potential disorder in a flat 
band system? Does disorder promote the localization effects of the flat band or does it aid in to mitigate the Anderson theorem by delocalizing the electrons,  
i. e. an inverse Anderson effect? These questions are particularly pertinent for 2D systems where short range fluctuations are dominant. Inverse Anderson effect 
was discussed in the context of non interacting 3D diamond lattice containing flat bands, wherein it was shown based on level statistics analysis that a weak disorder can destroy the phase coherence of the flat band energy states leading to delocalization and formation of extended states \cite{goda_prl2006}. The effect of the interplay of disorder and interaction was studied on a diamond chain comprising of three flat bands. It was demonstrated that as a function of increasing disorder the system undergoes transition between three different localizations. The CLS observed in the clean limit gives way to a flat-band localized state at weak disorder and eventually to Anderson localization at strong disorder \cite{roy_prr2020}. In a similar spirit, keeping in focus the quasi one dimensional Creutz lattice with attractive Hubbard interaction it was demonstrated that superconductivity is resilient to on-site disorder with the critical disorder strength for the SIT being proportional to the superfluid stiffness in the clean limit. The results suggest that flat band superconductivity is largely immune to potential disorder \cite{chan_arxiv2025}. Similar conclusions were drawn for metal/flat band material/metal heterostructure with the flat band material being of Lieb lattice geometry. It was shown that disorder promotes the electronic transport in the flat band material in contrast to the conventional materials with dispersive bands, an observation intimately tied to the quantum geometry of the flat band systems \cite{chau_arxiv2024}. More recently, based on BdG calculations on a Lieb lattice it was demonstrated that the flat band superconductivity is significantly robust as compared to the conventional superconductivity, particularly for off-diagonal disorder the suppression in T$_{c}$ with disorder is found to be quadratic in case of the flat band as compared to the linear suppression in conventional superconductors \cite{bouzerar_prbl2025}. Figure 8(c) shows the BKT transition temperature T$_{BKT}$ at the lattice 
filling of $\nu=1$ corresponding to the Fermi level present inside the dispersive band and at $\nu=3$ when the Fermi level is at flat band. It is observed that 
the flat band superconductivity is significantly robust as compared to the one originating from the dispersive band leading to both a higher T$_{BKT}$ and 
a larger disorder regime over which the superconductivity survives \cite{bouzerar_prbl2025}. A beyond mean field analysis of the interplay between disorder 
and interaction for a flat band superconductor is still awaited. Moreover, the nature of the disorder induced MIT in flat band materials is hitherto unexplored and promises to be intriguing particularly in systems such as, Kagome metal wherein transient localization of itinerant fermions via flat 
band induced local moments have been reported recently \cite{shashi_kagome2025,checkelsky_natphys2024}. 
\begin{figure*}
\begin{center}
\includegraphics[height=7.0cm,width=14cm,angle=0]{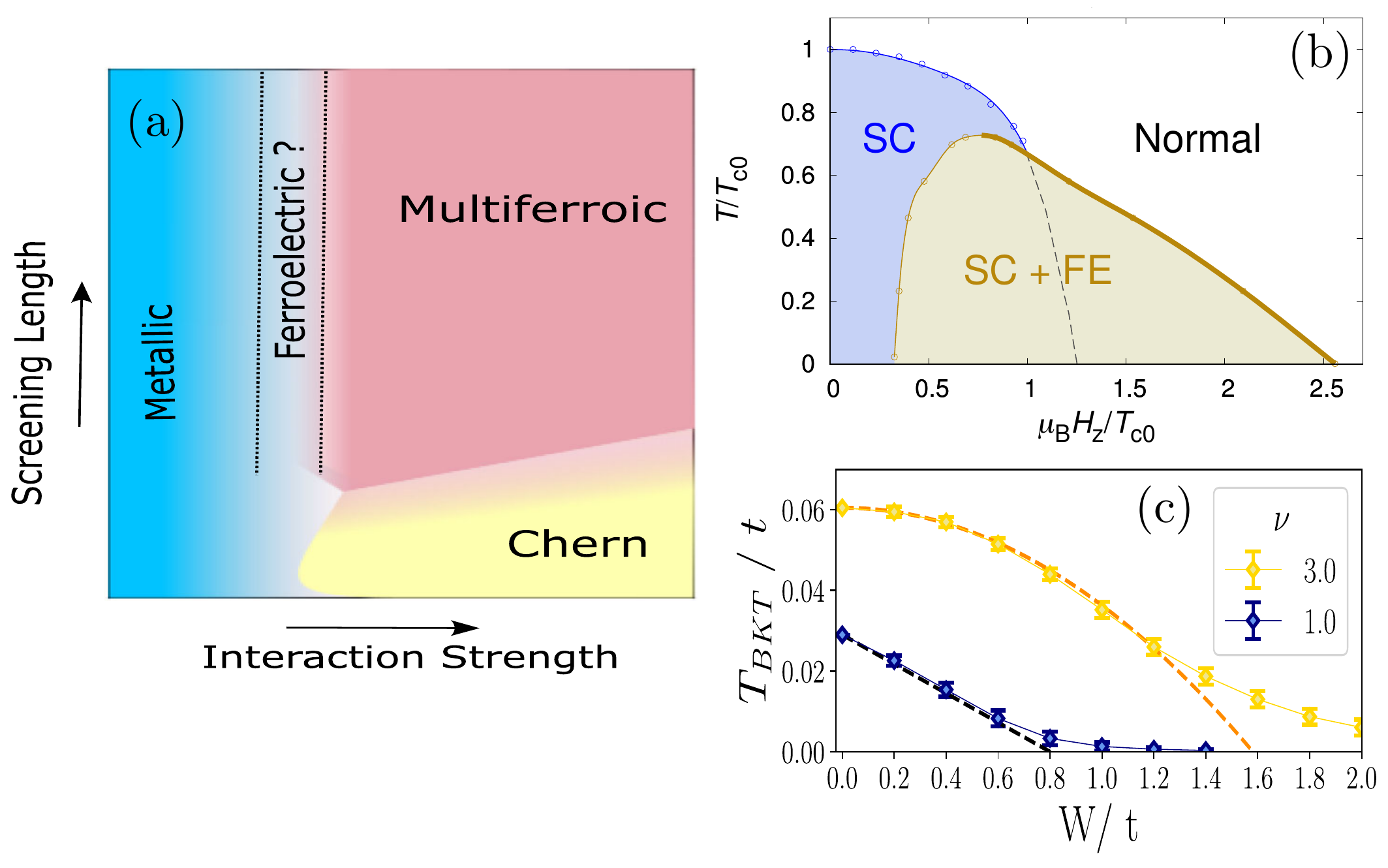}
\label{fig8}
\caption{(a) Multiferroic: schematic phase diagram for twisted TMD bilayers showing stable multiferroic phases. Figure adapted with permission from \cite{abouelkomsan_prl2024}. Copyright 2024, American Physical Society. (b) Ferroelectric superconductivity: Zeeman field ($h_{z}$) vs temperature 
($T$) phase diagram for a spin-orbit coupled attractive Hubbard model on a Lieb lattice geometry \cite{kanasugi_prb20218}. The spin-orbit coupling depicts the polar distortion of the lattice which accounts for the ferroelectric order. Note the coexisting superconducting and ferroelectric phases corresponding to ferroelectric superconductivity. Figure adapted with permission from \cite{kanasugi_prb20218}. Copyright 2018, American Physical Society. (c) Interplay 
of disorder and interaction: BKT temperature corresponding to potential disorder induced superconductor insulator transition (SIT) on a Lieb lattice 
at the lattice filling of $\nu=1$, corresponding to the Fermi level located inside the dispersive band and $\nu=3$, corresponding to the half filled case, 
obtained based on BdGMFT. Note the robustness of the superconducting pairing against disorder when the Fermi level is inside the flat band ($\nu=3$) 
\cite{bouzerar_prbl2025}. Figure adapted with permission from \cite{bouzerar_prbl2025}. Copyright 2025, American Physical Society.}
\end{center}
\end{figure*}

\subsection{Equilibrium and non equilibrium properties}
Ultrafast spectroscopic techniques—including time-resolved angle-resolved photoemission spectroscopy (tr-ARPES), resonant inelastic X-ray scattering (RIXS), and more recently, pump-probe experiments —enable selective probing, control, and manipulation of the various degrees of freedom in quantum many-body systems. These experiments primarily aim to investigate light-matter interactions in quantum materials, with the goal of stabilizing competing correlations as transient states. Such non-thermal pathways have opened new possibilities for realizing exotic quantum phases and their associated low-energy excitations, which have no counterparts in equilibrium. Efforts to understand the light matter interaction in flat band materials is recent and largely focussed on 
Kagome metals which owing to its frustrated geometry comprises of an energy landscape with competing correlations.  trARPES on CsV$_{3}$Sb$_{5}$ revealed interesting dynamics of the underlying CDW state and non thermal melting of the CDW order as well as its ultrafast recovery. The observed time 
scales and the associated collective modes suggest strong electron-phonon coupling as the origin of the CDW order in CsV$_{3}$Sb$_{5}$ \cite{azoury_pnas2023}. Similar electron-phonon coupling has been observed in Fe$_{3}$Sn$_{2}$, another Kagome metal,  via temperature and fluence dependent transient reflectivity measurements. Based on the dynamics of the charge carriers and coherent phonon modes in this material three different 
time scales are brought forth, while the fast and slow time scales could be explained to be tied to the metallic nature of this material the medium time scale 
could be understood only by attributing it to unconventional charge carriers, indicating the rich phase competition in this flat band system \cite{goncalvesfaria_npjquantmat2024}. Understanding the non equilibrium dynamics of the flat band materials and the possible dynamic stabilization of 
transient phases via light matter coupling is currently in vogue and is expected to be pursued extensively in the near future, both in terms of experiments 
and theory alike. 

\section{Acknowledgements}
 The author would like to acknowledge the support from the Anusandhan National Research Foundation, Govt. of India through the grant 
 ANRF CRG/2023/002593.
 
 \bibliography{references.bib}

\end{document}